\documentclass[preprint,aps,a4paper]{revtex4-1}

\usepackage{mathrsfs}
\usepackage{graphicx}
\usepackage{multirow}
\usepackage{makecell}
\usepackage{booktabs}
\usepackage{lineno}

\begin{document}
    \title{Ubiquitous Coexisting Electron-Mode Couplings in High Temperature Cuprate Superconductors}	
      
	
	
	\author{Hongtao Yan$^{1,2}$, Jin Mo Bok$^{3}$, Junfeng He$^{1\dag}$, Wentao Zhang$^{1\ddag}$, Qiang Gao$^{1}$, Xiangyu Luo$^{1,2}$, Yongqing Cai$^{1}$, Yingying Peng$^{1\S}$, Jianqiao Meng$^{1\P}$, Cong Li$^{1}$, Hao Chen$^{1,2}$, Chunyao Song$^{1,2}$, Chaohui Yin$^{1,2}$, Taimin Miao$^{1,2}$, Genda Gu$^{4}$, Chengtian Lin$^{5}$, Fengfeng Zhang$^{6}$, Feng Yang$^{6}$, Shenjin Zhang$^{6}$, Qinjun Peng$^{6}$, Guodong Liu$^{1,2,7}$, Lin Zhao$^{1,2,7}$, Han-Yong Choi$^{8}$, Zuyan Xu$^{6}$ and X. J. Zhou$^{1,2,7,9,*}$}
		
	\affiliation{
		\\$^{1}$National Lab for Superconductivity, Beijing National laboratory for Condensed Matter Physics, Institute of Physics, Chinese Academy of Sciences, Beijing 100190, China
		\\$^{2}$University of Chinese Academy of Sciences, Beijing 100049, China
		\\$^{3}$Department of Physics, Pohang University of Science and Technology (POSTECH), Pohang 37673, Korea
		\\$^{4}$Condensed Matter Physics, Materials Science Division of Brookhaven National Laboratory, Upton, NY 11973-5000, USA
		\\$^{5}$Max Planck Institute for Solid State Research, Heisenbergstrasse 1, D-70569 Stuttgart, Germany
		\\$^{6}$Technical Institute of Physics and Chemistry, Chinese Academy of Sciences, Beijing 100190, China
		\\$^{7}$Songshan Lake Materials Laboratory, Dongguan, Guangdong 523808, China
		\\$^{8}$Department of Physics, Sungkyunkwan University, Suwon 16419, Korea
		\\$^{9}$Beijing Academy of Quantum Information Sciences, Beijing 100193, China
		\\$^{\dag}$Present address: Department of Physics and CAS Key Laboratory of Strongly-Coupled Quantum Matter Physics, University of Science and Technology of China, Hefei, Anhui 230026, China
		\\$^{\ddag}$Present address: Key Laboratory of Artificial Structures and Quantum Control (Ministry of Education), Shenyang National Laboratory for Materials Science, School of Physics and Astronomy, Shanghai Jiao Tong University, Shanghai, China
		\\$^{\S}$Present address: International Center for Quantum Materials, School of Physics, Peking University, Beijing 100871, China
		\\$^{\P}$Present address: School of Physics and Electronics, Central South University, Changsha, Hunan 410083, China
		\\$^{*}$Corresponding author: XJZhou@iphy.ac.cn
	}
	
	\date{\today}
	
	\maketitle
	
	\newpage
	
	{\bf In conventional superconductors, the electron-phonon coupling plays a dominant role in pairing the electrons and generating superconductivity. In high temperature cuprate superconductors, the existence of the electron coupling with phonons and other boson modes and its role in producing high temperature superconductivity remain unclear. The evidence of the electron-boson coupling mainly comes from the angle-resolved photoemission (ARPES) observations of the $\sim$70\,meV nodal dispersion kink and the $\sim$40\,meV antinodal kink. However, the reported results are sporadic and the nature of the involved bosons are still under debate. Here we report new findings of ubiquitous two coexisting electron-mode couplings in cuprate superconductors. By taking ultra-high resolution laser-based ARPES measurements, combined with the improved second derivative analysis method, we discovered that the electrons are coupled simultaneously with two sharp phonon modes with energies of $\sim$70\,meV and $\sim$40\,meV in different superconductors with different doping levels, over the entire momentum space and at different temperatures above and below the superconducting transition temperature. The observed electron-phonon couplings are unusual because the associated energy scales do not exhibit an obvious change across the superconducting transition. We further find that the well-known ``peak-dip-hump" structure, which has long been considered as a hallmark of superconductivity, is also omnipresent and consists of finer structures that originates from electron coupling with two sharp phonon modes. These comprehensive results provide a unified picture to reconcile all the reported observations and pinpoint the origin of the electron-mode couplings in cuprate superconductors. They provide key information to understand the role of the electron-phonon coupling in generating high temperature superconductivity. 
	}

	\vspace{3mm}
	
	High temperature cuprate superconductors are derived from doping the parent Mott insulators\cite{YTokura1998MImada,XGWen2006PALee}. They exhibit anomalous normal state properties and unconventional superconductivity which are attributed to strong electron correlation and electron interactions with other collective excitations (boson modes) like phonons, magnetic fluctuations and so on\cite{BStatt1999TTimusk,JZaanen2015BKeimer}. Revealing such many-body effects is crucial to understanding the unusual properties and superconductivity mechanism in cuprate superconductors\cite{PWAnderson2007PWAnderson}. With the dramatic improvement of the instrumental resolutions, angle-resolved photoemission spectroscopy (ARPES) has emerged as a powerful technique to probe the many body effects in cuprate superconductors\cite{ZXShen2003ADamascelli,JRSchrieffer2006JRSchrieffer,WTZhang2018XJZhou,ZXShen2021JASobota}. It has been found that the band dispersion along the nodal direction exhibits a kink at $\sim$70\,meV in various cuprate superconductors although its origin remains under debate\cite{ZXShen2000PVBogdanov,MRNorman2000MEschrig,DGHinks2001PDJohnson,KKadowaki2001AKaminski,ZXShen2001ALanzara,ZXShen2003XJZhou,YAndo2006AAKordyuk,XJZhou2008WTZhang2,XJZhou2013JFHe}. Another dispersion kink at $\sim$40\,meV is also revealed near the antinodal region which is attributed to electron coupling with either phonons or magnetic resonance mode\cite{DSDessau2003ADGromko,JFink2003TKKim,KKadowaki2003TSato,ZXShen2004TCuk,NNagaosa2004TPDevereaux,XJZhou2013JFHe}. The well-known peak-dip-hump structure observed in the superconducting state near the antinodal region is considered as a hallmark of the electron-mode coupling in cuprate superconductors\cite{HRaffy1999JCCampuzano,KKadowaki2001AKaminski,ZXShen2018YHe}. It has been shown that superconductivity is closely related to the strength of the electron-mode couplings\cite{ZXShen2018YHe,GDGu2020TValla}. However, the observations of the electron-boson couplings in cuprate superconductors have been mostly sporadic and sometimes controversial (Fig. S1), lacking the consistency to pin down their nature and their role in generating high temperature superconductivity.
	 	
	In this paper, we report new findings and new understanding of the electron-mode couplings from our comprehensive high resolution laser-based ARPES measurements of the Bi-based cuprate superconductors. We discovered that the $\sim$40\,meV mode-coupling, which was observed only in Bi$_{2}$Sr$_{2}$CaCu$_{2}$O$_{8+\delta}$ (Bi2212) near the antinodal region, is also present in Bi$_{2}$Sr$_{2}$CuO$_{6+\delta}$ (Bi2201); it is also observed off the antinodal region and even along the nodal direction. On the other hand, the $\sim$70\,meV mode-coupling, which was observed mainly near the nodal area, is clearly revealed near the antinodal region. We have found the ubiquitous coexistence of the $\sim$40\,meV and $\sim$70\,meV mode-couplings in different cuprate superconductors over the entire momentum space, both in the normal and superconducting states and at different doping levels. The well-known ``peak-dip-hump" structure, which is usually observed in the photoemission spectra of Bi2212 near the antinodal region in the superconducting state, is actually composed of a finer ``peak-double dip-double hump" structure that originates from electron coupling with two sharp modes. These observations provide a unified picture to understand all the previous results and the nature of the electron-boson couplings in high temperature cuprate superconductors.
	
	ARPES measurements were performed using our lab-based laser ARPES systems equipped with the 6.994\,eV vacuum-ultra-violet (VUV) laser and a hemispherical electron energy analyzer or a time-of-flight electron energy analyzer (ARToF) with the capability of simultaneous two-dimensional momentum space detection\cite{XJZhou2008GDLiu,WTZhang2018XJZhou}. The energy resolution was set at $\sim$1\,meV and the angular resolution is $\sim$0.3\,$^\circ$, corresponding to a momentum resolution of $\sim$0.004\,\AA$^{-1}$ at the photon energy of 6.994\,eV. High quality single crystals of Bi2201, Bi2212 and Bi$_{2}$Sr$_{2}$Ca$_{2}$Cu$_{3}$O$_{10+\delta}$ (Bi2223) were grown by the traveling solvent floating zone method and post annealed in different atmospheres to get different doping levels\cite{XJZhou2009JQMeng2,XJZhou2010LZhao,XJZhou2012SYLiu,XJZhou2016YXZhang2,CTLin2016AMaljuk,XJZhou2019YDing} (Fig. S2 and Fig. S3). For convenience, we define hereafter the name of the samples according to their doping level and superconducting transition temperature ($T_{c}$) by using UD for the underdoped, OP for the optimally-doped and OD for the overdoped. All the samples were cleaved $\mathit{in\,situ}$ at a low temperature and measured in vacuum with a base pressure better than 5$\times$10$^{-11}$\,Torr. The Fermi level is referenced by measuring on a clean polycrystalline gold that is electrically connected to the sample.    
	
	Figure 1 shows the observation of two coexisting mode-couplings in the band structures measured in Bi2201 and Bi2212 along the nodal direction. Fig. 1a1 shows the band structure of an overdoped Bi2212 OD73K sample measured along the nodal direction at 17\,K. In order to reveal the fine electronic structures, we took second derivative of the original data with respect to energy to get images in Fig. 1b1 and Fig. 1c1. The detailed analysis of the band is presented in Fig. S4 in Supplementary Materials. The comparison between the original data (Fig. S4a) and the second derivative images (Fig. S4e and Fig. S4f), together with the comparison between the original photoemission spectra (energy distribution curves, EDCs, Fig. S4g) and their corresponding second derivative EDCs (Fig. S4h), indicates that the utilization of the second derivative method emphasizes on the curvature change, thus suppressing the dramatic intensity variation in the original data and sharpening the subtle features. This makes it possible to reveal fine structures in the second derivative images (Fig. S4e and Fig. S4f, Fig. 1b1 and Fig. 1c1) that are hard to see in the original data (Fig. S4a and Fig. 1a1). For this reason, such an improved analysis method will be used throughout the paper. The second derivative images are presented in two different ways with one to highlight the peak and hump features (Fig. 1b1) and the other to highlight the dip features (Fig. 1c1). As we will see below, the identification of both the hump structure and the dip structure are necessary to understand the electron-boson couplings in cuprate superconductors. 
	
	As seen in the second derivative image in Fig. 1b1, in addition to the main band that is clear in the original data (Fig. 1a1), four new features can be observed below the Fermi level. They show up as two flat bands on the left side of the main band, LLH (left-low energy-hump) and LHH (left-high energy-hump), and the other two flat bands on the right side of the main band, RLH (right-low energy-hump) and RHH (right-high energy-hump). In the other second derivative image in Fig. 1c1, four new features can also be observed below the Fermi level marked as LLD (left-low energy-dip), LHD (left-high energy-dip), RLD (right-low energy-dip) and RHD (right-high energy-dip). As illustrated from a typical second derivative EDC in Fig. 1d1, the new features in Fig. 1b1 represent the hump structures in EDCs (marked as LH and HH in Fig. 1d1) while in Fig. 1c1 they correspond to the dip structures in EDCs (marked as LD and HD in Fig. 1d1). These results indicate that, in addition to the main peak, coexisting double hump and double dip structures are observed in EDCs over a wide range of momentum space. These can also be directly observed from the original EDCs (Fig. S4g) and the corresponding second derivative EDCs (Fig. S4h). The energy position of the two hump and two dip structures in EDCs does not change when the momentum moves away from the main band and they have the same energy position on both the left and right sides of the main band (Fig. S4h). The energy position of the low energy hump (LH) and the high energy hump (HH) in Fig. 1(b1,d1) is $\sim$60\,meV and $\sim$91\,meV, respectively, while the energy position of the low energy dip (LD) and the high energy dip (HD) in Fig. 1(c1,d1) is $\sim$38\,meV and $\sim$75\,meV, respectively. 
	
	We emphasize that the high resolution and high statistics data from our laser ARPES measurements are critical in revealing the fine electronic structures in Bi2212 (Fig. 1(a1-d1)) that were not observed before. We also took high resolution laser ARPES measurements on the nodal band structure of the heavily overdoped non-superconducting Bi2201 (ODNSC) sample (Fig. 1a2). The observed results (Fig. 1a2-1d2) are rather similar to those in Bi2212 (Fig. 1a1-1d1). Four new features can be clearly observed below the Fermi level in the second derivative images (Fig. 1b2 and Fig. 1c2). In addition to the main peak, coexisting double hump (LH: $\sim$52\,meV and HH: $\sim$85\,meV in Fig. 1(b2,d2)) and double dip (LD: $\sim$34\,meV and HD: $\sim$65\,meV in Fig. 1(c2,d2)) structures are observed in Bi2201 in EDCs over a wide range of momentum space.
	
	The peculiar fine structures in the measured nodal band structures of Bi2212 (Fig. 1a1-1d1) and Bi2201 (Fig. 1a2-1d2) can be understood by considering electron coupling with two sharp boson modes. Fig. 1a3 shows the simulated single-particle spectral function by involving electron-boson coupling with two modes at 70\,meV and 40\,meV in the normal state (simulation details are described in Supplementary Materials). The corresponding second derivative images are shown in Fig. 1b3 and Fig. 1c3. It is clear that the simulated results (Fig. 1a3-1d3) show remarkable resemblance to the measured results in Bi2212 (Fig. 1a1-1d1) and Bi2201 (Fig. 1a2-1d2). These demonstrate unambiguously that in the nodal electron structures of Bi2212 and Bi2201 electrons are coupled with two sharp boson modes. In order to understand the measured data, several points need to be noted. First, as shown in Fig. S5 in Supplementary Materials, two sharp boson modes must be included to produce two hump structures (Fig. S5g) and two dip structures (Fig. S5h); one mode can only produce one hump structure (Fig. S5b) and one dip structure (Fig. S5c) that are not consistent with the measured results. Second, as shown in Fig. S5 and Fig. S6, the associated boson mode must have a finite linewidth which causes an energy separation between the hump and dip structures in EDCs. The boson mode with zero linewidth will result in the same energy position of the hump and dip structures in EDCs that are apparently not consistent with the measured results. Third, as shown in Fig. S6, when the involved boson modes have finite linewidths, the energy position of the dips in EDCs does not represent the mode energy. Instead, the mode energy is determined by the energy positions of both the hump and dip structures as a group, close to the middle position between them (Fig. S6d and Fig. S6h). In this way, the two energy scales in Bi2212 can be determined to be $\sim$83\,meV and $\sim$49\,meV from Fig. 1a1-1d1, and they are $\sim$75\,meV and $\sim$43\,meV in Bi2201 as determined from Fig. 1a2-1d2.
	
	The electron-boson coupling along the nodal direction has been studied extensively in Bi2212 and Bi2201 through the electron self-energy analysis and the $\sim$70\,meV dispersion kink is ubiquitously observed \cite{ZXShen2000PVBogdanov,MRNorman2000MEschrig,DGHinks2001PDJohnson,KKadowaki2001AKaminski,ZXShen2001ALanzara,ZXShen2003XJZhou,YAndo2006AAKordyuk,XJZhou2008WTZhang2,XJZhou2013JFHe}. We also carried out such an analysis on the high quality laser ARPES data of Bi2212 (Fig. 1a1) and Bi2201 (Fig. 1a2). Fig. 1e1 shows the nodal dispersion of Bi2212 obtained from Fig. 1a1 by fitting the momentum distribution curves (MDCs) at different binding energies. To reveal the kink structures in the dispersion, several straight lines are assumed as bare bands and the resultant ``effective real part of electron self-energy" is presented in Fig. 1f1. Two kinks are observed at $\sim$73\,meV and $\sim$39\,meV although the energy position slightly depends on the bare band selection. The observation of these two kinks in dispersion is consistent with the finding of the electron coupling with two modes in Bi2212 as shown in Fig. 1a1-1d1. From a similar self-energy analysis of the Bi2201 data in Fig. 1a2, two kinks can be identified at $\sim$73\,meV and $\sim$41\,meV (Fig. 1f2). It is consistent with the observation of the electron coupling with two modes in Bi2201 as shown in Fig. 1a2-1d2. We also carried out self-energy analysis on the simulated data in Fig. 1a3. The two dispersion kinks at $\sim$70\,meV and $\sim$40\,meV are in a good agreement with the electron coupling with two modes shown in Fig. 1a3-1d3. Overall, we have found clear evidence of electron coupling with two coexisting sharp modes in the nodal electronic structure of Bi2212 and Bi2201.
   
   Now we come to check on the electron coupling near the antinodal region using our second derivative analysis method. Fig. 2a1 shows a high resolution laser ARPES measured band of Bi2212 (OD73K) near the antinodal region; the corresponding second derivative images displayed in two different ways are shown in Fig. 2b1 and Fig. 2c1. Fig. 2d1 shows two EDCs at two typical momentum points; their corresponding second derivative EDCs are shown in Fig. 2e1. As seen in Fig. 2b1, in addition to the main band, two flat features can be observed below the bottom of the main band. These two flat features in Fig. 2b1 correspond to the two hump structures in EDCs at different momentum points, marked as LH and HH in Fig. 2d1 and Fig. 2e1 (detailed analysis can be found in Fig. S7 in Supplementary Materials). Likewise, the two flat features in Fig. 2c1 represent the two dip structures in EDCs, marked as LD and HD in Fig. 2d1 and Fig. 2e1. Fig. 2a2-2e2 show the measured band structure of Bi2201 (ODNSC) near the antinodal region (Fig. 2a2), its corresponding second derivative images (Fig. 2b2 and Fig. 2c2), typical EDCs (Fig. 2d2) and the corresponding second derivative EDCs (Fig. 2e2). Overall, the observed fine structures in Bi2201 are similar to those observed in Bi2212. Two flat features can also be observed in the second derivative images (Fig. 2b2 and Fig. 2c2) although they are relatively weaker than those in Bi2212 (Fig. 2b1 and Fig. 2c1). They also correspond to the double hump and double dip structures observed in EDCs (Fig. 2d2) and second derivative EDCs (Fig. 2e2). 
   
   The unusual fine structures in the measured band structures of Bi2212 (Fig. 2a1-2d1) and Bi2201 (Fig. 2a2-2d2) near the antinodal region can also be understood by considering electron coupling with two sharp boson modes. Fig. 2a3 shows the simulated band structure by considering electron coupling with two boson modes in the normal state. In the simulation, the energy position of the original main band bottom (35\,meV) is smaller than the energies of the two modes (70\,meV and 40\,meV). As seen in Fig. 2a3-2e3, the simulated results can well reproduce the observed band structures in Bi2212 and Bi2201, including the observation of two flat features in the second derivative images (Fig. 2b3 and Fig. 2c3) and the double hump and double dip structures in EDCs (Fig. 2d3) and second derivative EDCs (Fig. 2e3). These demonstrate unambiguously that in Bi2212 and Bi2201 electrons near the antinodal region also couple to two coexisting sharp boson modes. The two energy scales, determined from both the hump and dip positions, are $\sim$50\,meV and $\sim$90\,meV for Bi2212 (Fig. 2b1-2e1) and $\sim$40\,meV and $\sim$75\,meV for Bi2201 (Fig. 2b2-2e2). 
   
   We note that, when the bandwidth is significantly larger than the mode energy, the electron-mode coupling produces kink in the dispersion near the mode energy as demonstrated in the simulated data of Fig. 1(a3,e3,f3) and observed in Bi2212 (Fig. 1(a1,e1,f1)) and Bi2201 (Fig. 1(a2,e2,f2)). On the other hand, when the bandwidth is comparable or smaller than the mode energy, the electron-mode coupling can no longer produce kink in the dispersion. Instead, the electron-boson coupling causes band renormalization; the bandwidth shrinks from the original 35\,meV to 23\,meV as shown in the simulated data of Fig. 2a3. The above results in Fig. 1 and Fig. 2 have demonstrated the advantages of the second derivative method in identifying electron-mode coupling. When the bandwidth is larger than the mode energy, although both the band dispersion analysis and the second derivative analysis can be applied to identify the mode-coupling and the results from these two methods are consistent, the second derivative analysis is more reliable than the self-energy analysis because the flat features in the second derivative images are more characteristic and robust than the subtle slope change in the electron self-energy. In particular, when the bandwidth is comparable or smaller than the mode energy, the traditional self-energy analysis is no longer useful and the second derivative analysis provides a more general and decisive way to pin down the electron-mode coupling.
   
   The electron-mode coupling has been observed in Bi2212 near the antinodal region, famously known as the $\sim$40\,meV antinodal mode-coupling that is identified through the band dispersion and electron self-energy analysis\cite{DSDessau2003ADGromko,JFink2003TKKim,KKadowaki2003TSato,ZXShen2004TCuk,NNagaosa2004TPDevereaux,XJZhou2013JFHe}. Through the band dispersion analysis, we observed a similar $\sim$40\,meV energy scale in Bi2212 near the antinodal region on both the bonding band and the antibonding band, as shown in Fig. S8a. The measured results and related analyses in Fig. 2 provide significant new insights on the electron-mode coupling in cuprates superconductors. First, the $\sim$40\,meV antinodal mode-coupling has been observed only in Bi2212 so far; we have observed, for the first time, the $\sim$40\,meV antinodal mode-coupling in Bi2201 even in a heavily overdoped non-superconducting sample (Fig. 2a2-2e2). Second, we have clearly observed the coexisting $\sim$40\,meV and $\sim$70\,meV energy scales near the antinodal region in Bi2212 and, for the first time, also in Bi2201. The dispersion kink at $\sim$70\,meV was once observed in optimally-doped Bi2212 near the antinodal region in the superconducting state and was interpreted as due to the $\sim$40\,meV mode-coupling with its energy position shifted by the superconducting gap\cite{DSDessau2003ADGromko,ZXShen2004TCuk}. Our simultaneous observation of the $\sim$40\,meV and $\sim$70\,meV energy scales near the antinodal region in heavily overdoped non-superconducting Bi2201 indicates that these two mode-couplings are separate and coexisting. 
    
    Figure 3 shows the momentum evolution of the two mode-couplings between the nodal and the antinodal regions in Bi2212 and Bi2201. Fig. 3a shows the band structures of Bi2212 OD73K sample measured along different momentum cuts; here the second derivative images are presented to highlight the hump structures caused by the electron-boson couplings. The two mode-couplings, observed along the nodal direction (Fig. 1a1-1f1 and Fig. 3a1) and near the antinodal region (Fig. 2a1-2e1 and Fig. 3a6), are present in the entire measured momentum space as evidenced by the observation of two coexisting hump structures marked as LH and HH in Fig. 3a1-3a6. Fig. 3b shows the second derivative EDCs from the corresponding band structures in Fig. 3a1-3a6. Here the EDCs on the right side of the Fermi momentum are chosen because as shown in the simulated data in Fig. S5, when the momentum point moves away from the dispersion region, the hump and dip structures become more pronounced. Two hump structures (LH and HH) and two dip structures (LD and HD) can be clearly observed in each EDC in Fig. 3b. From the energy positions of the hump and dip structures in Fig. 3b, the energy scales of the two mode-couplings are determined and plotted in Fig. 3c. When the momentum cuts change from the nodal direction to the near antinodal region, the two energy scales show little change with momentum within the uncertainty of our data measurements and analysis. The coexistence of the two mode-couplings is also observed in Bi2201 over the entire measured momentum space although the signal of the low energy $\sim$40\,meV energy scales is relatively weak (Fig. 3e1-3e6 and Fig. 3f). These two energy scales also exhibit little change with momentum variation from the nodal direction to the antinodal region (Fig. 3g). We also determined the energy scales of the two mode-couplings by using the traditional self-energy method in the momentum space where such a method is applicable. Fig. 3i shows the measured electron self-energy of Bi2212 along six momentum cuts (the details of the self-energy extraction can be found in Supplementary Materials and Fig. S9). Two energy scales are observed and the low energy scale (LScale) becomes pronounced with the momentum cuts changing from the nodal direction to the near antinodal region. The two energy scales observed (marked as LScale and HScale in Fig. 3i) are plotted in Fig. 3c; they show a good agreement with the results obtained from the second derivative method. Fig. 3j shows the electron self-energy of Bi2201 measured along different momentum cuts from the nodal direction to the antinodal regions (the details of the self-energy extraction can be found in Supplementary Materials). Two main energy scales can be observed with the low energy scale (LScale) getting relatively pronounced and becoming dominant when the momentum cuts change from the nodal direction to the near antinodal region. The two energy scales determined from the self-energy in Fig. 3j are plotted in Fig. 3g; they are also consistent with those results obtained from the second derivative analysis.        
    
    We have observed the coexisting two mode-couplings in different superconductors with various doping levels. As seen from Fig. S10 in Supplementary Materials, the two mode-couplings can be observed from the band structures along the nodal direction in Bi2201 over a wide doping range (Fig. S10a), in underdoped, optimally-doped and overdoped Bi2212 (Fig. S10b) and in Bi2223 (Fig. S10c). Note that these two mode-couplings are clearly observable even in heavily overdoped non-superconducting Bi2201 and heavily overdoped Bi2212 with a $T_{c}$ of 35\,K ($p$$\sim$0.25). Our observation of the two mode-couplings over the entire doping range and even in the heavily overdoped non-superconducting Bi2201 sample is distinct from the previous reports that the antinodal low energy mode-coupling becomes significantly weakened upon crossing the critical doping of $\sim$0.19\cite{ZXShen2018YHe} or completely disappears when the Bi2212 sample is heavily-overdoped to be  non-superconducting\cite{GDGu2020TValla}. The energy scales of the two mode-couplings, within the experimental uncertainty, fall into two categories: the high energy scale lies in the energy range of (79$\pm$10)\,meV and the low energy scale lies in the energy range of (46$\pm$10)\,meV (Fig. S10f). The two mode-couplings are ubiquitous; they can be observed in Bi2201, Bi2212 and Bi2223 superconductors, at different doping levels and in the entire momentum space (Fig. 3, Fig. S10 and Fig. S11). These results indicate that the coexisting two mode-couplings is a common phenomenon in cuprate superconductors. 
    
    In order to understand the nature of the two coexisting mode-couplings, it is crucial to investigate their temperature dependence, particularly the change across the superconducting transition. To this end, we have taken high resolution laser ARPES data on Bi2212 and Bi2201 at different temperatures, as shown in Fig. 4. Fig. 4a shows the band structures of the Bi2212 OD73\,K sample measured along the nodal direction at different temperatures. The corresponding second derivative EDCs are plotted in Fig. 4d. The high energy mode-coupling can be clearly observed at 100\,K and the low energy one is visible at 90\,K; both are above the $T_{c}$ of 73\,K of the measured sample. Fig. 4b shows the band structures of the Bi2212 OD73\,K sample measured near the antinodal region at different temperatures. The corresponding second derivative EDCs are plotted in Fig. 4e. Both the high and the low energy mode-couplings can be observed up to 100\,K that is above the $T_{c}$ of 73\,K of the measured sample. Fig. 4c shows the band structures of the Bi2201 OD17\,K sample measured at different temperatures. The corresponding second derivative EDCs are plotted in Fig. 4f. The high energy mode-coupling can be clearly observed at 100\,K and the low energy one is visible at 80\,K; both are well above the $T_{c}$ of 17\,K of the measured sample. These results indicate that both mode-couplings exist above $T_{c}$. The pseudogap temperature $T^{*}$ of the Bi2212 OD73\,K sample is $\sim$89\,K\cite{SShin2015TKondo} and it is $\sim$30\,K for the Bi2201 OD17K sample\cite{CTLin2005GQZheng}. These further indicate that the two mode-couplings are present above $T^{*}$. In fact, the observation of the two energy scales even in the heavily overdoped non-superconducting Bi2201 (Fig. 1a2-1f2, Fig. 2a2-2e2 and Fig. 3(e1-e6,f,g)) already demonstrates their presence in the normal state that has nothing to do with the superconducting transition or the pseudogap transition. Our observation on the persistence of the low energy $\sim$40\,meV energy scale in both Bi2212 and Bi2201 above the temperatures of $T_{c}$ and $T^{*}$ is distinct from the previous reports that the $\sim$40\,meV energy scale observed in the superconducting state of Bi2212 near the antinodal region disappears above the superconducting transition in terms of the self-energy analysis\cite{DSDessau2003ADGromko,KKadowaki2003TSato,KKadowaki2006KTerashima2,XJZhou2013JFHe,GDGu2020TValla}. We note that the temperature dependence of the two mode-couplings observed in the Bi2212 OD73\,K sample along the nodal direction (Fig. 4a and Fig. 4d) is quite similar to that observed in the Bi2201 OD17\,K sample (Fig. 4c and Fig. 4f). This strongly indicates that the weakening of the mode-coupling signatures with the increasing temperature in Bi2212 (Fig. 4d) and Bi2201 (Fig. 4f) is mainly due to thermal broadening rather than superconducting transition.
    
    The energy scales of the two mode-couplings do not exhibit an obvious change across the superconducting transition or pseudogap temperature. Fig. 4g shows the temperature dependence of the two energy scales in the band structures along the nodal direction for the Bi2212 OD73K sample. The two energy scales are obtained from both the second derivative EDCs in Fig. 4d (solid circles in Fig. 4g) and from the self-energy analyses in Fig. 4j for the high energy scale and in Fig. 4k for the low energy scale (empty circles in Fig. 4g). The results obtained from these two methods are consistent with each other. Both the energy scales show little change with temperature; there is no obvious change at either the $T_{c}$ of 73\,K or the pseudogap temperature $T^{*}$ of $\sim$89\,K. Fig. 4h shows the temperature dependence of the two energy scales in the band structures near the antinodal region for the Bi2212 OD73K sample obtained from the second derivative EDCs in Fig. 4e (solid circles in Fig. 4h) and from the self-energy analysis in Fig. 4l for the low energy scale (empty circles in Fig. 4h). Again, both the energy scales keep nearly constant with temperature and no obvious change at the $T_{c}$ or the pseudogap temperature $T^{*}$ is observed. Fig. 4i shows the temperature dependence of the two energy scales for the Bi2201 OD17K sample obtained from the second derivative EDCs in Fig. 4f (solid circles in Fig. 4i). Both the energy scales also show little change with temperature. There is no obvious change at either the $T_{c}$ of 17\,K or the pseudogap temperature $T^{*}$ of $\sim$30\,K. We note that, even in two different superconductors Bi2212 and Bi2201 with quite different $T_{c}$s of 73\,K and 17\,K and different superconducting gap sizes of $\sim$20\,meV and $\sim$8\,meV, the observed two coexisting energy scales ($\sim$80\,meV and $\sim$40\,meV) and their temperature dependence are quite similar (Fig. 4g-4i). These results rule out the possibility that, in the previous reports, the observed $\sim$70\,meV kink near nodal\cite{ZXShen2008WSLee} or antinodal regions\cite{ZXShen2004TCuk,ZXShen2018YHe} and the $\sim$40\,meV kink near the antinodal region\cite{ZXShen2004TCuk} in the superconducting state of Bi2212 are considered to originate from a low energy mode shifted by the superconducting gap.
    
    Our new findings of the ubiquitous two coexisting sharp mode-couplings in different materials, in the entire momentum space, at different doping levels and at different temperatures have put strong constraints on the nature of the associated electron-boson coupling. The similar energy scales and the similar momentum-, doping- and temperature-dependences observed in Bi2201, Bi2212 and Bi2223 indicate that the two mode-couplings have a common origin. As the energy scales exhibit a weak temperature dependence and do not show abrupt change across $T_{c}$ and $T^{*}$, the energy of the two modes can be determined with one being $\sim$70\,meV and the other being $\sim$40\,meV although the exact values may vary slightly with temperature, doping or in different materials. The previous observations of the nodal $\sim$70\,meV mode-coupling\cite{ZXShen2000PVBogdanov,MRNorman2000MEschrig,DGHinks2001PDJohnson,KKadowaki2001AKaminski,ZXShen2001ALanzara,ZXShen2003XJZhou,YAndo2006AAKordyuk,XJZhou2008WTZhang2,XJZhou2013JFHe} and the antinodal $\sim$40\,meV mode-coupling in the superconducting state Bi2212\cite{DSDessau2003ADGromko,JFink2003TKKim,KKadowaki2003TSato,ZXShen2004TCuk,NNagaosa2004TPDevereaux,XJZhou2013JFHe} are consistent with our present results but cover only a portion of the global electron-boson coupling landscape (Fig. S1). The origin of the $\sim$70\,meV nodal kink remains unsettled with possible mechanisms like electron-phonon coupling\cite{ZXShen2001ALanzara,MKrauss2003KPBohnen,SGLouie2008FGiustino,DManske2008RHeid,TPDevereaux2008DReznik,NNagaosa2004TPDevereaux,SGLouie2021ZLLi}, electron coupling with magnetic resonance mode or spin fluctuation\cite{DGHinks2001PDJohnson,BKeimer2009TDahm}, or purely electron-electron correlation\cite{DVollhardt2007KByczuk}. The nature of the $\sim$40\,meV mode in the antinodal kink observed in the superconducting state of Bi2212 is also under debate between possible candidates of phonons\cite{ZXShen2004TCuk,NNagaosa2004TPDevereaux} or magnetic resonance mode\cite{DSDessau2003ADGromko,GDGu2020TValla}. Our observations of the $\sim$70\,meV mode-coupling in the broader material, momentum, doping and temperature spaces lend further support that the involved boson most likely corresponds to the phonon of the oxygen breathing mode\cite{YEndoh1999RJMcQueeney}. We have found that the $\sim$40\,meV mode-coupling is no longer limited to the antinodal region of Bi2212 in the superconducting state; it is also present in other momentum space, above the superconducting transition temperature and even in heavily overdoped non-superconducting Bi2201 sample. These results rule out the possibility of the magnetic resonance mode as the origin of the $\sim$40\,meV mode are strongly in favor of the phonon of the oxygen buckling mode\cite{IAAksay1995DReznik,AZawadowski1999TPDevereaux}. 
    
    The momentum and temperature dependences of the identified $\sim$70\,meV and $\sim$40\,meV mode-couplings are quite unusual in that their energy scales do not change with momentum and temperature in the $d$-wave superconductors. In a conventional picture where the electron-boson coupling vertex is momentum independent, an Einstein phonon mode with an energy $\mathrm{\Omega}$ in the normal state is expected to be shifted to $\mathrm{\Omega}$+$\mathrm{\Delta_{0}}$ ($\mathrm{\Delta_{0}}$ is the maximum of the $d$-wave superconducting gap) in the superconducting state over the entire Fermi surface\cite{NEBickers2004AWSandvik}. We have carried out similar simulations by considering phonon modes with a finite linewidth (Fig. S12 and Fig. S13 in Supplementary Materials) and found the same conclusion. Therefore, such a conventional electron-phonon coupling picture can not explain the temperature independence of the two mode-couplings we observed. In an alternative electron-phonon coupling picture where the dominant forward scattering is considered, a phonon mode with an energy $\mathrm{\Omega}$ in the normal state is expected to be shifted to $\mathrm{\Omega}$+$\mathrm{\Delta_{\mathbf{k}}}$ ($\mathrm{\Delta_{\mathbf{k}}}$ is the size of the local $d$-wave superconducting gap) in the superconducting state along the Fermi surface\cite{OVDolgov2005MLKulic,TPDevereaux2012SJohnston}. This picture can explain the zero energy shift of the modes across $T_{c}$ along the nodal direction where the local gap is zero. But it can not explain the absence of energy shift for the two modes crossing $T_{c}$ over the entire momentum space when the local gap increases from zero at nodal to the maximum $\mathrm{\Delta_{0}}$ at the antinodal region. Further theoretical efforts are needed to understand the unusual momentum and temperature dependences of the two electron-phonon couplings we have observed. 
    
    Our discovery of the two coexisting sharp mode-couplings that is omnipresent in cuprate superconductors has significant implications on the origin of the well-known peak-dip-hump structure observed in the superconducting state of Bi2212 near the ($\mathit{\pi}$,0) antinodal region\cite{AKapitulnik1991DSDessau,NKoshizuka1999AVFedorov,HRaffy1999JCCampuzano}. Since it was observed only near the antinodal region, it was proposed that the peak-dip-hump structure originates from electron coupling with collective excitations with a $\mathbf{Q}$=($\mathit{\pi}$,$\mathit{\pi}$) like spin fluctuations or magnetic resonance mode\cite{JRSchrieffer1997ZXShen,KKadowaki1997MRNorman,DKMorr1998AVChubukov,AVChubukov1999AAbanov,MRNorman2000MEschrig}. This peak-dip-hump structure is considered as a hallmark of superconductivity and is used to study the relationship between the electron-boson coupling and superconductivity\cite{AJMillis2000JOrenstein,ZXShen2018YHe}. Our present work has provided fundamentally new insights on the nature of the peak-dip-hump structure in cuprate superconductors. First, as shown in Fig. 5, the peak-dip-hump structure actually consists of one peak, two humps (LH and HH) and two dips (LD and HD), forming a peak-double dip-double hump structure. This comes directly from the two coexisting mode-couplings we have observed. Second, different from the previous observations of the peak-dip-hump structure only near the antinodal region, the peak-double dip-double hump structure is observed over the entire momentum space, as shown in Fig. S4, Fig. S7 and Fig. 5. It can even be observed near the nodal region (Fig. 5(a1,c1) and Fig. S14). Third, unlike the previous observation of the peak-dip-hump structure only in Bi2212, the peak-double dip-double hump structure can also be observed in Bi2201 as shown in Fig. 5(c1-c2,d1-d3)  and Fig. S15. Fourth, different from the previous observation of the peak-dip-hump structure only in the superconducting state, the peak-double dip-double hump structure can also be observed in the normal state as shown in Fig. S15. It can even be clearly observed in the heavily overdoped non-superconducting Bi2201 (Fig. 5(c1-c2,d1-d3)). These observations strongly indicate that the peak-double dip-double hump structure is also ubiquitous. It originates from electron coupling with two sharp phonon modes other than the spin fluctuation or magnetic resonance mode as proposed before\cite{JRSchrieffer1997ZXShen,KKadowaki1997MRNorman,DKMorr1998AVChubukov,AVChubukov1999AAbanov,MRNorman2000MEschrig}. 
    
    Our present work has provided new and comprehensive information on the electron-phonon coupling in cuprate superconductors. The observation of the two coexisting sharp mode-couplings provides a unified picture to reconcile the previous results of the nodal kink and antinodal kink in cuprate superconductors. In conventional superconductors, the electron-phonon coupling is always present at different temperatures both above and below $T_{c}$. This does not mean that the electron-phonon coupling is not relevant to superconductivity. Instead, it has been well established that it is the electron-phonon coupling that induces the electron pairing and the superconductivity\cite{JRSchrieffer1957JBardeen}. Now we have found the ubiquitous existence of the electron-coupling simultaneously with two sharp phonon modes at $\sim$70\,meV and $\sim$40\,meV in different cuprate superconductors, at different doping levels and at different temperatures above and below $T_{c}$. We have also shown that the observed electron-phonon couplings are unusual because their temperature and momentum dependences can not be understood in the conventional electron-phonon coupling picture. This may be related to the fact that the cuprate superconductors are doped Mott insulators with strong electron-electron correlation\cite{XGWen2006PALee}. Our results ask for further theoretical efforts to understand the role of the observed electron-phonon couplings in generating high temperature superconductivity in cuprate superconductors. 
    
    \vspace{3mm}
    

    \vspace{3mm}
    
    \noindent {\bf Acknowledgement} This work is supported by the National Natural Science Foundation of China (Grant Nos. 11888101, 11922414 and 11974404), the National Key Research and Development Program of China (Grant Nos. 2016YFA0300300, 2016YFA0300602, 2017YFA0302900, 2018YFA0305602 and 2018YFA0704200), the Strategic Priority Research Program (B) of the Chinese Academy of Sciences (Grant No. XDB25000000 and XDB33000000), the Youth Innovation Promotion Association of CAS (Grant No. 2017013), the Research Program of Beijing Academy of Quantum Information Sciences (Grant No. Y18G06). J.M.B. was supported by National Research Foundation (NRF) of Korea through Grants  No. NRF-2019R1I1A1A01057393. The work at BNL was supported by the US Department of Energy, oﬃce of Basic Energy Sciences, contract No. DOE-sc0012704. H.Y.C. was supported by the National Research Foundation of Korea under NRF-2021R1F1A1063697.

    \vspace{3mm}
    
    \noindent {\bf Author Contributions}\\
    X.J.Z. and H.T.Y. proposed and designed the research. H.T.Y., J.F.H., W.T.Z., Q.G., X.Y.L., Y.Y.P., J.Q.M. and L.Z. carried out the ARPES experiments. H.T.Y., Y.Y.P., J.Q.M., G.D.G., C.T.L. and L.Z. grew the single crystals. H.T.Y., J.F.H., W.T.Z., Q.G., X.Y.L., Y.Q.C., Y.Y.P., J.Q.M., C.L., H.C., C.Y.S., C.H.Y., T.M.M., F.F.Z., F.Y., S.J.Z., Q.J.P., G.D.L., L.Z., Z.Y.X. and X.J.Z. contributed to the development and maintenance of Laser-ARPES systems. J.M.B. and H.Y.C. contributed to theoretical analysis. H.T.Y and X.J.Z. analyzed the data and wrote the paper. All authors participated in discussions and comments on the paper.
       
    \vspace{3mm}
    
    \noindent {\bf\large Additional information}\\
    \noindent{\bf Competing financial interests:} The authors declare no competing financial interests.

    \newpage
    
    \begin{figure*}[tpb]
    \begin{center}
    	\includegraphics[width=1.0\columnwidth,angle=0]{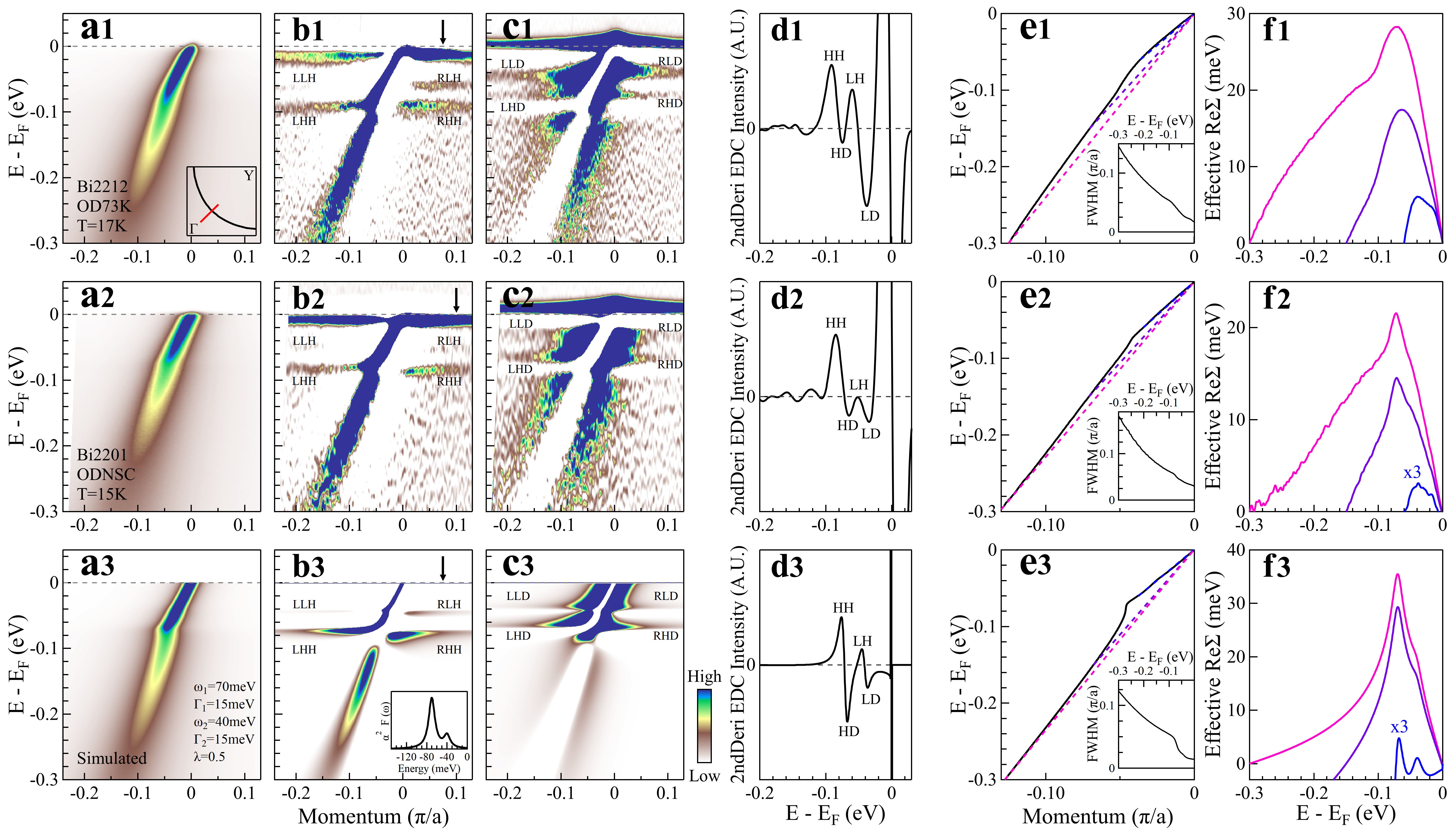}
    \end{center}
    
    \caption{{\bf Observation of two coexisting mode-couplings in the band structures along the nodal direction.} (a1) Band structure of the Bi2212 OD73K sample taken at 17\,K along the nodal direction. The location of the momentum cut is shown in the bottom-right inset by the red line. (b1) Second derivative image of (a1) with respect to energy showing the absolute magnitude of the negative values. (c1) Second derivative image of (a1) with respect to energy showing the absolute magnitude of the positive values. (d1) Second derivative EDC at the momentum point marked by the arrow in (b1). For convenience, we show the negative of the second derivative values throughout the paper. In this case, the peaks correspond to the observed band structure in (a1). (e1) Band dispersion obtained by fitting MDCs at different energies from (a1). The pink, purple and blue dashed straight lines connecting two energy positions in the dispersion at the Fermi level and -0.3\,eV, -0.15\,eV and -0.06\,eV, respectively, represent empirical bare bands to extract effective real parts of electron self-energy in (f1). The inset shows the corresponding MDC width (Full width at half maximum, FWHM). (f1) Effective real parts of electron self-energy from the dispersion in (e1) by subtracting different empirical bare bands. (a2-f2) Same as (a1-f1) but for the Bi2201 ODNSC sample measured at 15\,K. (a3-f3) Same as (a1-f1) but for the simulated band structure. Two modes, 40\,meV and 70\,meV, are assumed in the simulation, as shown in the inset of (b3) (the details of the simulation are described in Supplementary Materials). In (b1,b2,b3), in addition to the main band, four features appear marked as LLH, RLH, LHH and RHH. In (c1,c2,c3), four features are observed marked as LLD, RLD, LHD and RHD. In (d1,d2,d3), the observed features are marked as HH, LH, HD and LD. As shown in Fig. S4 in Supplementary Materials, the hump and dip features in (b1,b2,b3) and (c1,c2,c3) correspond to the hump and dip structures in the corresponding EDCs and second derivative EDCs.}
   
    \end{figure*}

    \begin{figure*}[tbp]
	\begin{center}
		\includegraphics[width=1.0\columnwidth,angle=0]{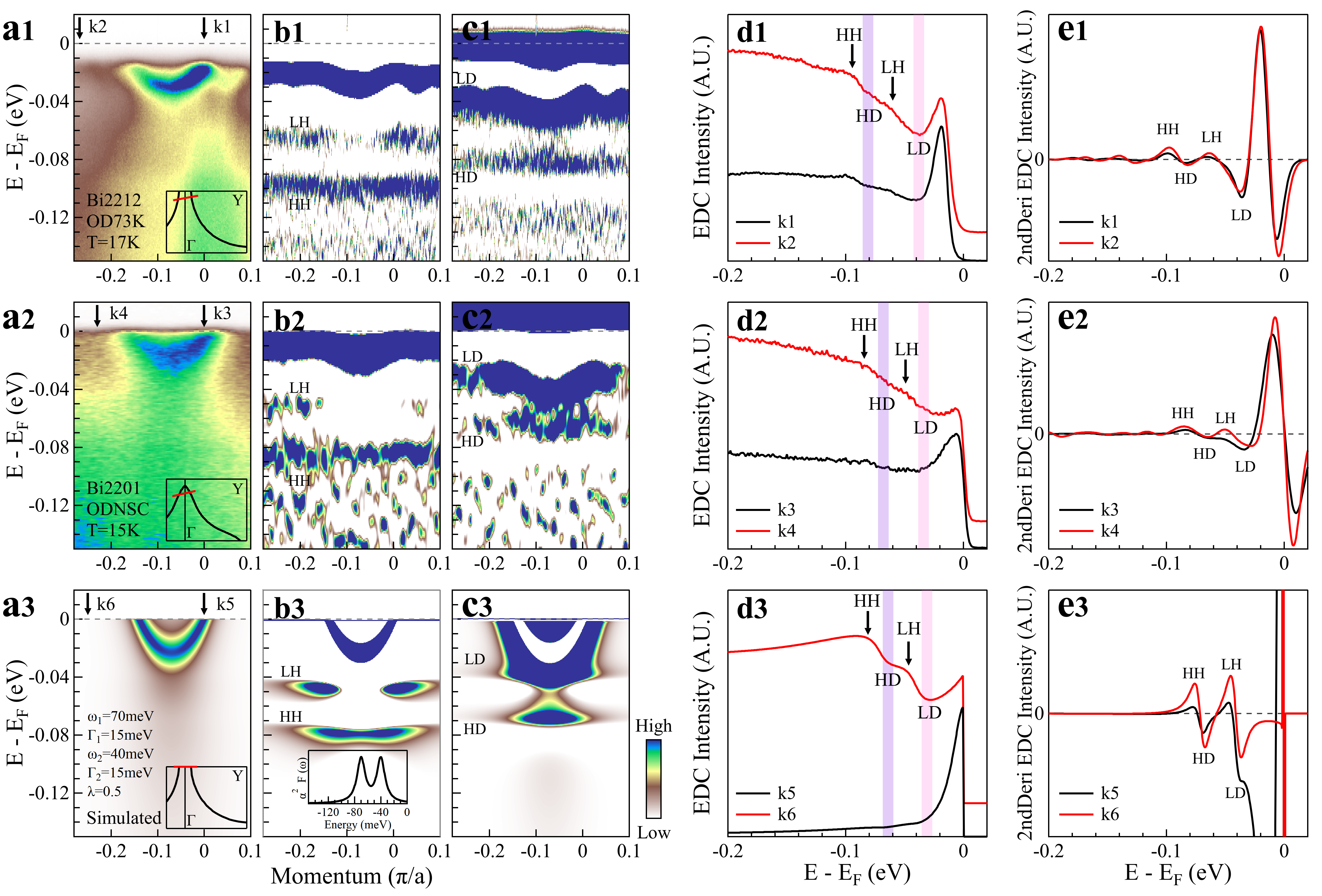}
	\end{center}
	
	 \caption{{\bf Observation of two coexisting mode-couplings in the band structures near the antinodal region.} (a1) Band structure of the Bi2212 OD73K sample taken at 17\,K near the antinodal region. The location of the momentum cut is shown in the bottom-right inset by the red line. (b1) Second derivative image of (a1) with respect to energy showing the absolute magnitude of the negative values. (c1) Second derivative image of (a1) with respect to energy showing the absolute magnitude of the positive values. (d1) EDCs at two momentum points, k1 and k2, as marked in (a1). For clarity, the data are normalized by the peak intensity near the Fermi level and are offset along the vertical axis. (e1) Second derivative EDCs obtained from (d1). For clarity, the data are normalized by the intensity difference between the HH and LD features. (a2-e2) Same as (a1-e1) but for the Bi2201 ODNSC sample measured at 15\,K. (a3-e3) Same as (a1-e1) but for the simulated band structure. Two modes, 40\,meV and 70\,meV, are assumed in the simulation, as shown in the inset of (b3) (the details of the simulation are described in Supplementary Materials). In (b1,b2,b3), in addition to the main band, two features appear marked as LH and HH. In (c1,c2,c3), two features are observed marked as LD and HD. In (d1,d2,d3), the observed features are marked as HH, LH, HD and LD. Correspondingly, in (e1,e2,e3), the observed features are marked as HH, LH, HD and LD. As shown in Fig. S7 in Supplementary Materials, the hump and dip features in (b1,b2,b3) and (c1,c2,c3) correspond to the hump and dip features in the measured EDCs.}

	\end{figure*}

    \begin{figure*}[tbp]
	\begin{center}
		\includegraphics[width=1.0\columnwidth,angle=0]{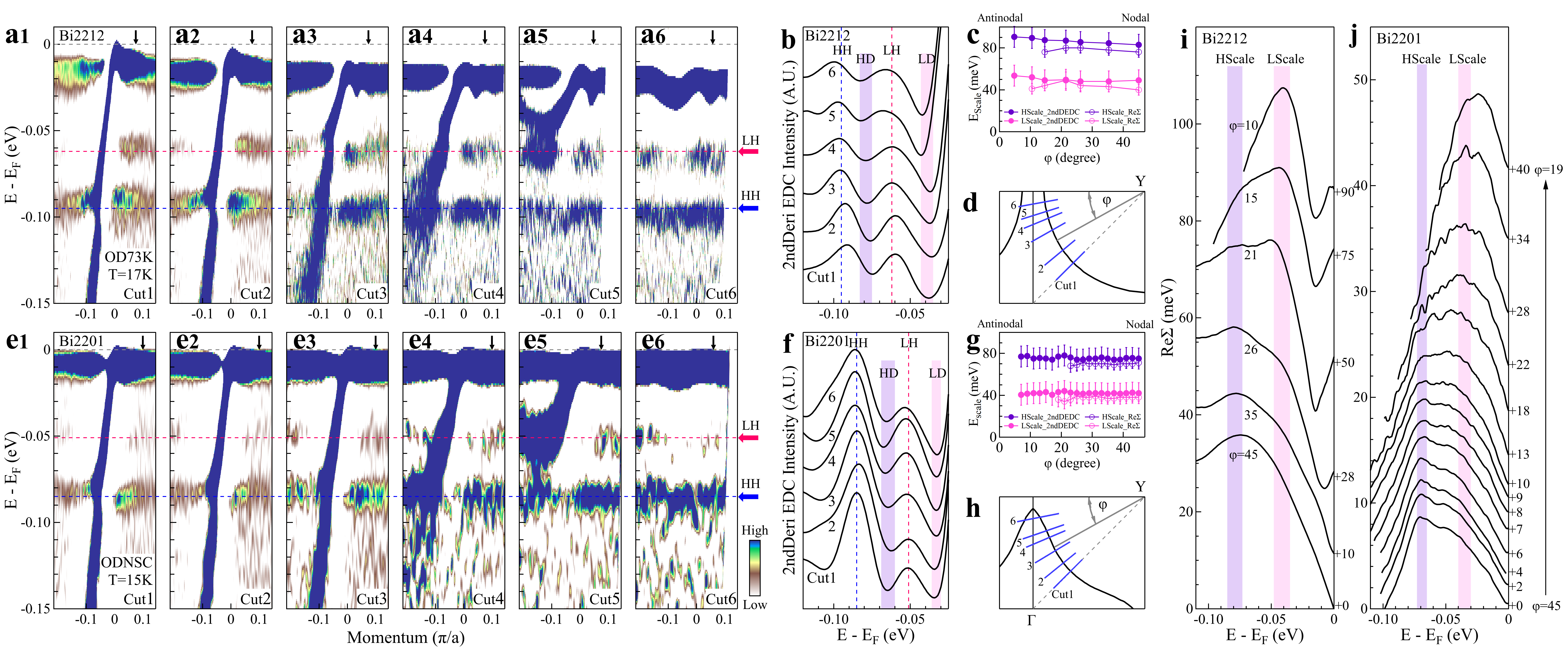}
	\end{center}
	
	\caption{{\bf Momentum dependence of the two mode-couplings in Bi2212 and Bi2201.} Details of the electron self-energy extraction are described in Supplementary Materials. (a) Momentum-dependent band structures of the Bi2212 OD73K sample taken at 17\,K. These are the second derivative images with respect to energy showing the absolute magnitude of the negative values. The location of the six momentum cuts are shown in (d) by blue lines. Two features are marked by LH (pink dashed line) and HH (blue dashed line). (b) Second derivative EDCs obtained from (a). The corresponding momentum position is marked by arrow in each panel of (a). For clarity, the data are normalized by the intensity difference between the HH and LD features, and are offset along the vertical axis. The two dip and two hump features in these second derivative EDCs are marked by LD, HD, LH and HH. (c) The two energy scales obtained from the second derivative EDCs in (b) (solid circles)  and the real part of electron self-energy in (i) (empty circles). From the second derivative EDCs in (b), the high energy scale (HScale) is determined from the middle point between HH and HD features while the low energy scale (LScale) is from the middle point between LH and LD features. (d) The location of the momentum cuts used in (a). (e-h) Same as (a-d) but for the Bi2201 ODNSC sample measured at 15\,K. (i) Real parts of the electron self-energy of the Bi2212 OD73K sample taken at 17\,K along different momentum cuts. Here the momentum cut positions are defined by the angle $\varphi$ as marked in (d). For clarity, the data are offset along the vertical axis with the offset values marked on the right side of each curve. Two energy scales can be observed as marked by HScale and LScale. Their positions are plotted also in (c) as empty circles. (j) Real parts of the electron self-energy of the Bi2201 ODNSC sample taken at 15\,K along different momentum cuts. The momentum cut positions are defined by the angle $\varphi$ marked in (h). They are taken every two degrees between $\varphi$=45\,$^\circ$ and $\varphi$=19\,$^\circ$. For clarity, the data are offset along the vertical axis with the offset values marked on the right side of each curve. Two energy scales can be seen as marked by HScale and LScale. Their positions are also plotted in (g) as empty circles.} 
	
	\end{figure*}

    \begin{figure*}[tbp]
	\begin{center}
		\includegraphics[width=1.0\columnwidth,angle=0]{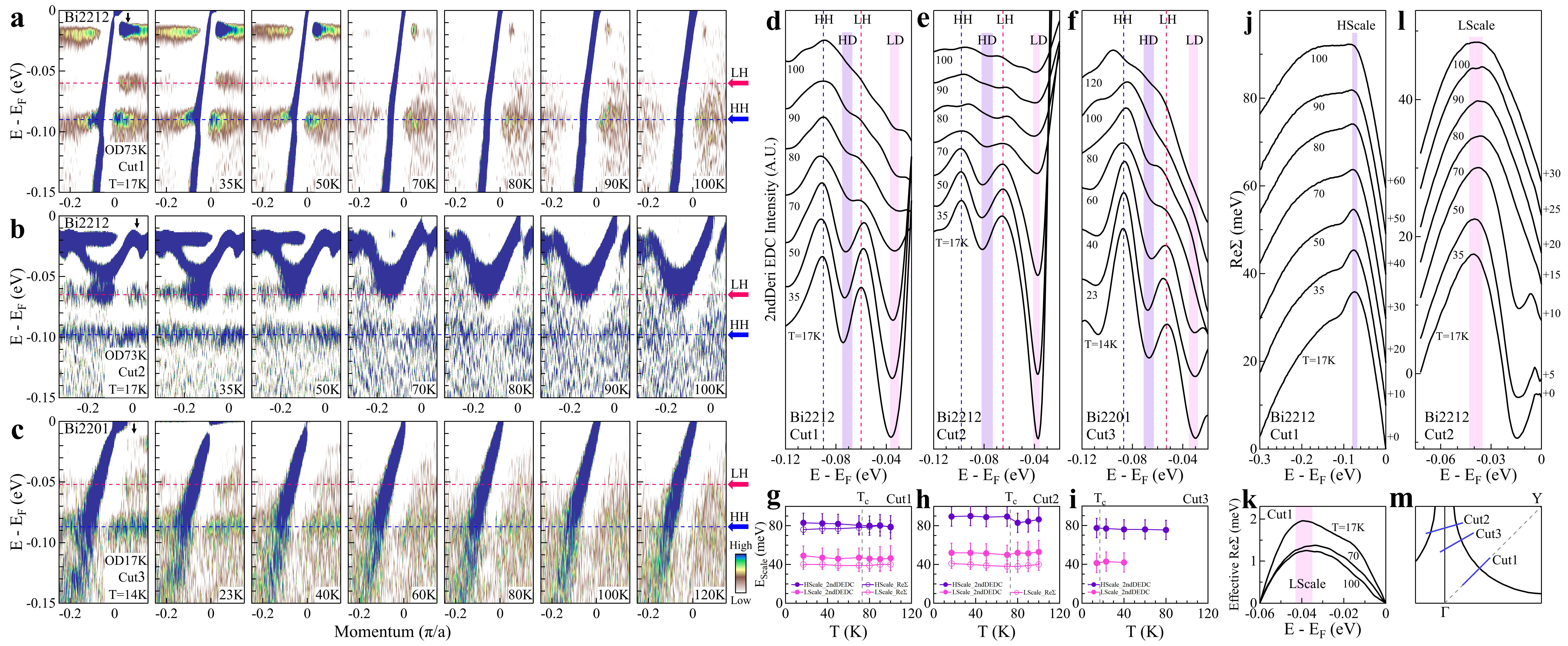}
	\end{center}
	
	\caption{{\bf Temperature dependence of the two mode-couplings in Bi2212 and Bi2201.} The locations of the momentum cuts are shown in (m) by blue lines. (a) Band structures of the Bi2212 OD73K sample taken along the nodal direction at different temperatures. These are the second derivative images with respect to energy showing the absolute magnitude of the negative values. The Fermi-Dirac distribution is removed from the images. Two features are marked by LH (pink dashed line) and HH (blue dashed line). (b) Same as (a) but for the momentum Cut 2 near the antinodal region. (c) Same as (a) but taken from Bi2201 OD17K sample along the momentum Cut 3. (d) Second derivative EDCs obtained from (a). The corresponding momentum position is marked by an arrow in the left-most panel of (a). For clarity, the data are offset along the vertical axis. The two dip and two hump features are marked by LD, HD, LH and HH, respectively. (e) Same as (d) but from (b). (f) Same as (d) but from (c). (g) The energy positions of the two modes obtained from the second derivative EDCs in (d) (solid circles)  and the real part of electron self-energy in (j) (empty circles). (h) Same as (g) but from (e) and (l). (i) Same as (g) but from (f). (j) Real parts of the electron self-energy of the band structure measured along the nodal directional Cut 1 at different temperatures. Details of the electron self-energy extraction are described in Supplementary Materials. For clarity, the data are offset along the vertical axis with the offset values marked on the right side of each curve. The energy scale, marked as HScale, can be clearly observed and its energy position is plotted in (g) as empty circles. (k) Effective real parts of the electron self-energy from dispersions taken along the nodal direction at different temperatures. They are obtained by subtracting the empirical bare bands that are straight lines connecting two energy positions in the dispersion at the Fermi level and -0.06\,eV. The energy scale, marked as LScale, can be observed and its energy position is also plotted in (h) as empty circles. (l) Same as (j) but for the momentum Cut 2 near the antinodal region. The energy scale, marked as LScale, can be observed and its energy position is included in (h) as empty circles.}

	\end{figure*}

    \begin{figure*}[tbp]
	\begin{center}
		\includegraphics[width=1.0\columnwidth,angle=0]{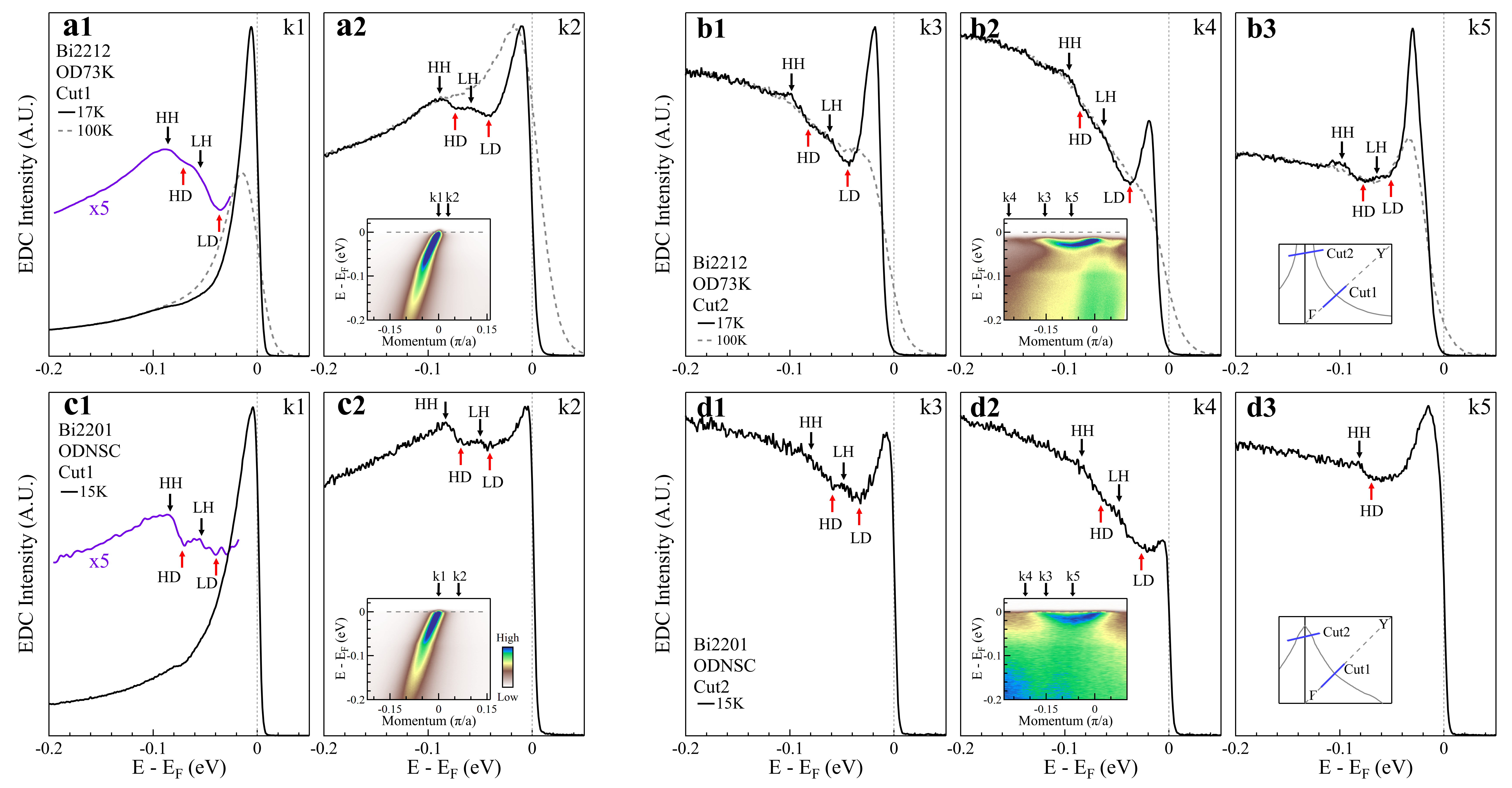}
	\end{center}

	\caption{{\bf Observation of the ``peak-double dips-double humps" structure in Bi2212 and Bi2201.} (a1-a2) EDCs of the Bi2212 OD73K sample measured along the nodal direction at two momentum points k1 (Fermi momentum, a1) and k2 (a2). The measured band structure is shown in the inset of (a2) where the location of the two momentum points k1 and k2 are marked by the arrows. The location of the momentum cut is shown in the inset of (b3) by a blue line marked as Cut 1. Two EDCs are shown in (a1) and (a2) that are measured at 17\,K (black curve) and 100\,K (gray dashed curve). The inset purple line in (a1) is the difference between the EDC at 17\,K and a fitted Lorentzian; for clarity, it is multiplied by five times ($\times$5) (the details to extract the purple line are described in Supplementary Materials Fig. S14c). From the purple line in (a1) and the EDC at 17\,K in (a2), two dip and two hump features can be observed marked as LD, HD, LH and HH, respectively. (b1-b3) EDCs of the Bi2212 OD73K sample taken along the momentum cut near the antinodal region at three momentum points k3 (Fermi momentum, b1), k4 (b2) and k5 (point at the (0,0)-(0,$\mathit{\pi}$) line, b3). The measured band structure is shown in the inset of (b2) where the location of the three momentum points k3, k4 and k5 are marked by the arrows. The location of the momentum cut is shown in the inset of (b3) by a blue line marked as Cut 2. Two EDCs are shown in (b1), (b2) and (b3) that are measured at 17\,K (black curve) and 100\,K (gray dashed curve). In (b1-b3), from the EDCs at 17\,K, two dip (LD and HD) and two hump (LH and HH) features can be observed and marked. (c1-c2) Same as (a1-a2) but for the EDCs measured in the Bi2201 ODNSC sample at 15\,K. (d1-d3) Same as (b1-b3) but for the EDCs measured in the Bi2201 ODNSC sample at 15\,K.}
	
	\end{figure*}
	
\end{document}


\title{{\it Supplementary Materials for}\\Ubiquitous Coexisting Electron-Mode Couplings in High Temperature Cuprate Superconductors}
	
	\author{Hongtao Yan$^{1,2}$, Jin Mo Bok$^{3}$, Junfeng He$^{1\dag}$, Wentao Zhang$^{1\ddag}$, Qiang Gao$^{1}$, Xiangyu Luo$^{1,2}$, Yongqing Cai$^{1}$, Yingying Peng$^{1\S}$, Jianqiao Meng$^{1\P}$, Cong Li$^{1}$, Hao Chen$^{1,2}$, Chunyao Song$^{1,2}$, Chaohui Yin$^{1,2}$, Taimin Miao$^{1,2}$, Genda Gu$^{4}$, Chengtian Lin$^{5}$, Fengfeng Zhang$^{6}$, Feng Yang$^{6}$, Shenjin Zhang$^{6}$, Qinjun Peng$^{6}$, Guodong Liu$^{1,2,7}$, Lin Zhao$^{1,2,7}$, Han-Yong Choi$^{8}$, Zuyan Xu$^{6}$ and X. J. Zhou$^{1,2,7,9,*}$}

    \affiliation{
	\\$^{1}$National Lab for Superconductivity, Beijing National laboratory for Condensed Matter Physics, Institute of Physics, Chinese Academy of Sciences, Beijing 100190, China
	\\$^{2}$University of Chinese Academy of Sciences, Beijing 100049, China
	\\$^{3}$Department of Physics, Pohang University of Science and Technology (POSTECH), Pohang 37673, Korea
	\\$^{4}$Condensed Matter Physics, Materials Science Division of Brookhaven National Laboratory, Upton, NY 11973-5000, USA
	\\$^{5}$Max Planck Institute for Solid State Research, Heisenbergstrasse 1, D-70569 Stuttgart, Germany
	\\$^{6}$Technical Institute of Physics and Chemistry, Chinese Academy of Sciences, Beijing 100190, China
	\\$^{7}$Songshan Lake Materials Laboratory, Dongguan, Guangdong 523808, China
	\\$^{8}$Department of Physics, Sungkyunkwan University, Suwon 16419, Korea
	\\$^{9}$Beijing Academy of Quantum Information Sciences, Beijing 100193, China
	\\$^{\dag}$Present address: Department of Physics and CAS Key Laboratory of Strongly-Coupled Quantum Matter Physics, University of Science and Technology of China, Hefei, Anhui 230026, China
	\\$^{\ddag}$Present address: Key Laboratory of Artificial Structures and Quantum Control (Ministry of Education), Shenyang National Laboratory for Materials Science, School of Physics and Astronomy, Shanghai Jiao Tong University, Shanghai, China
	\\$^{\S}$Present address: International Center for Quantum Materials, School of Physics, Peking University, Beijing 100871, China
	\\$^{\P}$Present address: School of Physics and Electronics, Central South University, Changsha, Hunan 410083, China
	\\$^{*}$Corresponding author: XJZhou@iphy.ac.cn
    }
	
	\date{\today}
	
	\maketitle
	
	\newpage
	
	\noindent{\large{\bfseries 1. Summarized ARPES observations of the electron-boson couplings in B2201, Bi2212 and Bi2223}	
		
	Figure S1 shows the summarized ARPES observations of the electron-boson couplings in Bi2201, Bi2212 and Bi2212 from the previous reports\cite{ZXShen2000PVBogdanov,DGHinks2001PDJohnson,KKadowaki2001AKaminski,ZXShen2001ALanzara,ZXShen2003XJZhou,DSDessau2003ADGromko,JFink2003TKKim,KKadowaki2003TSato,ZXShen2004TCuk,YAndo2006AAKordyuk,XJZhou2008WTZhang2,XJZhou2013JFHe,SShin2013TKondo} and our present work. \\

	\noindent{\large{\bfseries 2. Samples and doping levels}
		
	Figure S2 shows the magnetic measurement results of  the Bi2201, Bi2212 and Bi2223 single crystal samples used in the present ARPES measurements. High quality single crystals of these Bi-based cuprate superconductors were grown by the traveling solvent floating zone method\cite{XJZhou2009JQMeng2,XJZhou2010LZhao,XJZhou2012SYLiu,CTLin2016AMaljuk}. We changed the doping level of Bi2201 by varying the La content in Bi$_{2}$Sr$_{2-x}$La$_{x}$CuO$_{6+\delta}$ (La-Bi2201) or Pb concentration in Bi$_{2.12-x}$Pb$_{x}$Sr$_{1.88}$CuO$_{6+\delta}$ (Pb-Bi2201), combined with the post annealing of the samples in different atmospheres at different temperatures\cite{XJZhou2009JQMeng2,XJZhou2010LZhao,XJZhou2019YDing} (Fig. S2a). The doping level of the Bi2212 samples was changed by post annealing in different atmospheres at different temperatures\cite{XJZhou2016YXZhang2} (Fig. S2b). A heavily overdoped Bi2212 sample was prepared  $\mathit{in\,situ}$ in the ARPES chamber by annealing in ozone atmosphere (Fig. S3). The Bi2223 sample is obtained by annealing the as-grown sample in flowing air at 550$^\circ$C (Fig. S2c). The doping level of the underdoped and optimally-doped Bi2201 samples is estimated from the $T_{c}$$\sim$doping level correspondence established before\cite{TMurayama2000YAndo}. The doping level of the overdoped Bi2201 samples is determined directly from the Fermi surface area measured by ARPES\cite{XJZhou2019YDing}. The doping level of the Bi2212 samples is estimated from the empirical relation, $T_{c}=T_{c,max}[1-82.6(p-0.16)^{2}]$, with a $T_{c,max}$=91\,K\cite{MRPresland1991NEFlower}. \\

	\noindent{\large{\bfseries 3. Observation of the two coexisting mode-couplings in the band structures measured along the nodal direction in Bi2212}
		
	Figure S4 shows the observation of two coexisting mode-couplings in the band structures measured in Bi2212 along the nodal direction. Fig. S4a shows the band structure of the Bi2212 OD73K sample along the nodal direction at 17\,K. Fig. S4b shows a typical EDC from Fig. S4a, where a main peak (peak), a low energy hump (LH), a high energy hump (HH), a low energy dip (LD) and a high energy dip (HD) can be observed. Fig. S4c shows the second derivative of the EDC in Fig. S4b. The sign of the second derivative EDC in Fig. S4c is reversed to get the curve in Fig. S4d, and it shows a direct correspondence to the original EDC in Fig. S4b. In this way, the subtle features in the original data (Fig. S4b) become more pronounced in the second derivative EDC (Fig. S4d), and the peak (peak), dip (LD and HD) and hump features (LH and HH) in the second derivative data (Fig. S4d) correspond well to those in the original data (Fig. S4b). For this reason, we will use this method throughout the paper. Correspondingly, two types of second derivative images can be obtained. One is to show the absolute magnitude of the negative values of the second derivative EDC in Fig. S4c, as shown in Fig. S4e, which highlights the peak and hump features. The other is to show the absolute magnitude of the positive values of the second derivative EDC in Fig. S4c, as shown in Fig. S4f, which highlights the dip features. Four features can be observed in the second derivative image in Fig. S4e below the Fermi level marked by LLH (left-low energy-hump), LHH (left-high energy-hump), RLH (right-low energy-hump) and RHH (right-high energy-hump). Four features can also be observed in the other second derivative image in Fig. S4f below the Fermi level marked by LLD (left-low energy-dip), LHD (left-high energy-dip), RLD (right-low energy-dip) and RHD (right-high energy-dip). Fig. S4g shows the typical EDCs at different momentum points where two humps (LH and HH) and two dips (LD and HD) can be observed in each EDC. Fig. S4h shows the corresponding second derivative EDCs obtained from the EDCs in Fig. S4g where the two hump (LH and HH) and two dip (LD and HD) features can be more clearly observed. \\

	\noindent{\large{\bfseries 4. Simulations of the mode-coupling in the normal state}

	We simulated the single-particle spectral functions $A(\mathbf{k}, \omega)$ in the normal state at zero temperature\cite{GGrimvall1981GGrimvall}. The ARPES intensity $I(\mathbf{k}, \omega)$ is expressed as:
	\begin{equation}
		I(\mathbf{k}, \omega, T)=f(\omega, T)A(\mathbf{k}, \omega)
	\end{equation}
	\noindent where $f(\omega, T)$ is the Fermi-Dirac distribution function. The spectral function $A(\mathbf{k}, \omega)$ is expressed as:
	\begin{equation}
		A(\mathbf{k}, \omega)=-\frac{1}{\pi} \frac{\Sigma^{''}(\mathbf{k}, \omega)}{[\omega-\epsilon(\mathbf{k})-\Sigma^{'}(\mathbf{k}, \omega)]^{2}+[\Sigma^{''}(\mathbf{k}, \omega)]^{2}}
	\end{equation}
	\noindent in which $\epsilon(\mathbf{k})$ represents the bare band. The real part ($\Sigma^{'}(\mathbf{k}, \omega)$) and imaginary part ($\Sigma^{''}(\mathbf{k}, \omega)$) of the electron self-energy $\Sigma(\mathbf{k}, \omega)$ can be expressed as:
	\begin{equation}
		\Sigma^{'}(\mathbf{k}, \omega)=\int_{Es}^{0} d\Omega\, \alpha^{2}F(\Omega) ln\left|\frac{\Omega+\omega}{\Omega-\omega} \right|
	\end{equation}
	\begin{equation}
		\Sigma^{''}(\mathbf{k}, \omega)=\Sigma^{''}_{imp}+\beta\omega^{2}-\pi\int_{Es}^{0} d\Omega \,\alpha^{2}F(\Omega)
	\end{equation}
	\noindent Here $\alpha^{2}F(\Omega)$ is the Eliashberg function and $Es$ is the cutoff of the Eliashberg function. The first term of $\Sigma^{''}(\mathbf{k}, \omega)$ is from impurity scattering, the second term of $\Sigma^{''}(\mathbf{k}, \omega)$ is from the electron-electron interaction in the Fermi liquid picture and the third term comes from the electron-boson coupling. 
	
	The electron-boson coupling strength, $\lambda$, is defined as:
	\begin{equation}
		\lambda=-2\int_{Es}^{0}\frac{d\Omega}{\Omega} \,\alpha^{2}F(\Omega)
	\end{equation}
	
	In order to simulate the electron coupling with sharp modes, we assume a Lorentzian form of the Eliashberg function for each mode:
	\begin{equation}
		\alpha^{2}F(\Omega)=\frac{C(\Gamma/2)^{2}}{(\Omega+\omega_{0})^{2}+(\Gamma/2)^{2}}
	\end{equation}
    \noindent Here $\omega_{0}$ is the mode energy, $\Gamma$ is the linewidth of the Lorentzian peak (full width at half maximum, FWHM) and $C$ is a constant. 
    	
	We first simulated the spectral function of the electron coupling with one mode as shown in Fig. S5a. To make the simulation comparable to the band structure measured along the nodal direction in Bi2212 (Fig. 1a1), we take the bare band $\epsilon(k)=2.4\,k$, the coefficient of electron-electron interaction $\beta=-1$ and $\Sigma^{''}_{imp}$=0.017\,eV. The unit of $\epsilon(k)$, $\omega$ and $\Omega$ is $eV$, and the unit of $k$ is $\pi/a$, where $a$ is the distance of the two nearest Cu atoms in the $\mathrm{CuO_{2}}$ plane. The mode energy is 70\,meV, its linewidth is 15\,meV and the electron-boson coupling strength $\lambda=0.5$. Fig. S5b and S5c are the second derivative images of Fig. S5a in two different ways. Fig. S5d shows the EDCs obtained from the spectral function in Fig. S5a at three different momentum points. Their corresponding second derivative EDCs are shown in Fig. S5e. When only one mode is involved in the mode-coupling, only one dip and one hump structures can be produced in EDCs.	
	
	We then simulated the spectral function of the electron coupling with two modes as shown in Fig. S5f. Here the same simulation parameters are used except that two modes are considered in the simulation. The two mode energies are 70\,meV and 40\,meV, both the linewidths are 15\,meV and the electron-boson coupling strength $\lambda$ is 0.5. Fig. S5g and S5h are the second derivative images of Fig. S5f in two different ways. Fig. S5i shows the EDCs obtained from the spectral function in Fig. S5f at three different momentum points. Their corresponding second derivative EDCs are shown in Fig. S5j. When two sharp modes are involved in the mode-coupling, two dip and two hump features can be produced in EDCs.
	
	In order to find out the mode energy when the mode involved in the electron-boson coupling has a finite linewidth, we simulated the spectral functions by considering different mode linewidth for the single mode case (Fig. S6(a-d)) and two modes case (Fig. S6(e-h)). In the single mode case, the positions of the hump and dip change with the mode linewidth as shown in Fig. S6d. But the middle point of the hump and dip features is very close to the energy of the mode. These results indicate that, when the involved boson mode has a finite linewidth, the energy position of the dip in EDCs no longer represents the mode energy. It is the middle point between the dip and hump features that represents the mode energy. The same conclusion holds when two modes are involved in the electron-boson coupling as seen in Fig. S6h.
	
	When the bandwidth is comparable or smaller than the involved mode energy, the spectral function of the electron-boson coupling can still be simulated by the formula (1)-(6). To make the simulation comparable to the band structure measured near the antinodal region in Bi2212 (Fig. 2a1), we take the bare band $\epsilon(k)=-0.035+7.12(k-0.07011)^{2}$, the coefficient of electron-electron interaction $\beta=-1$ and $\Sigma^{''}_{imp}$=0.017\,eV. The two mode energies are 70\,meV and 40\,meV, both the linewidths are 15\,meV and the electron-boson coupling strength $\lambda$ is 0.5. The simulated spectral function is shown in Fig. 2a3. \\

	\noindent{\large{\bfseries 5. Observation of the two coexisting mode-couplings in the band structures measured near the antinodal region in Bi2212}
	
	Figure S7 shows the observation of two coexisting mode-couplings in the band structures measured in Bi2212 near the antinodal region. Fig. S7a shows the band structure of the Bi2212 OD73K sample near the antinodal region at 17\,K. Two types of second derivative images obtained from Fig. S7a are shown in Fig. S7b and Fig. S7c. Four features can be observed in the second derivative image in Fig. S7b marked by LLH, LHH, RLH and RHH. Two features can be observed in the other second derivative image in Fig. S7c marked by LHD and RHD. Fig. S7d shows the typical EDCs at different momentum points where two humps (LH and HH) and two dips (LD and HD) can be observed in each EDC. Fig. S7e shows the corresponding second derivative EDCs obtained from the EDCs in Fig. S7d where the two hump (LH and HH) and two dip (LD and HD) features can be more clearly observed. \\

     \noindent{\large{\bfseries 6. Observation of the $\sim$40\,meV kink in the band structure measured at 17\,K near the antinodal region in Bi2212}
     	
     Figure S8a shows the $\sim$40\,meV kink in the band structure of the Bi2212 OD73K sample measured at 17\,K near the antinodal region. 
     The bonding band (BB) and antibonding band (AB) can be observed simultaneously. The black line and red line are the dispersions for the right antibonding band (AB$\_$R) and left bonding band (BB$\_$L), respectively, obtained by fitting MDCs at different energies. A kink with the energy of $\sim$40\,meV can be observed clearly in both the dispersions. For comparison, we show the second derivative image of Fig. S8a in Fig. S8b and Fig. S8c. Four features marked as LLH, LHH, RLH and RHH can be observed in Fig. S8b which represent the hump features in EDCs, and two features marked as LHD and RHD can be seen in Fig. S8c which represent the dip features in EDCs. The dip and hump features in the second derivative images indicate that $\sim$40\,meV and $\sim$70\,meV mode-couplings coexist in the band structure. The $\sim$40\,meV kink in the band structure in Fig. S8a corresponds to the $\sim$40\,meV mode-coupling obtained from the second derivative analysis in Fig. S8b and Fig. S8c. \\

     \noindent{\large{\bfseries 7. Extraction of the real part of the electron self-energy}
    
    Figure S9 shows the extraction of the real part of the electron self-energy from the measured band structures. Fig. S9a and S9b show the band structures of the Bi2212 OD73K sample measured along the nodal direction and near the antinodal region, respectively, at 17\,K. The band dispersions obtained by fitting the MDCs at different energies from Fig. S9a and S9b are shown in Fig. S9c and S9d, respectively, by the black lines. In order to extract the real part of the electron self-energy, we take the bands from the band structure calculations\cite{ABansil2005RSMarkiewicz} as the bare bands, shown as the red lines in Fig. S9c and S9d. The extracted real parts of the electron self-energy from Fig. S9c and S9d are shown in Fig. S9e and S9f, respectively. \\

     \noindent{\large{\bfseries 8. Doping dependence of the two coexisting mode-couplings in Bi2201, Bi2212 and Bi2223}
     
     Figure S10 shows the doping dependence of the two mode-couplings in Bi2201, Bi2212 and Bi2223 measured along the nodal direction. Fig. S10a shows the nodal band structures of Bi2201 over a wide doping range from underdoped all the way to the heavily overdoped non-superconducting samples. Fig. S10b shows the nodal band structures of the underdoped, optimally-doped and overdoped Bi2212, including the heavily overdoped sample with a $T_{c}$ at $\sim$35\,K. Fig. S10c shows the nodal band structure of the Bi2223 sample with a $T_{c}$ of 108K. The corresponding second derivative EDCs obtained from the band structures in Fig. S10(a-c) are shown in Fig. S10e. The energy positions of the two energy scales obtained from the second derivative EDCs in Fig. S10e are summarized in Fig. S10f. The two coexisting mode-couplings are observed in all the measurements along the nodal direction in Bi2201, Bi2212 and Bi2223 with different doping levels. The energy scales of the two mode-couplings, within the experimental uncertainty, fall into two categories, the high energy one lies at (79$\pm$10)\,meV while the low energy one lies at (46$\pm$10)\,meV. 
     
     Figure S11 shows the doping dependence of the two mode-couplings in Bi2201 and Bi2212 near the antinodal region. 
     Fig. S11(a-c) show the band structures of Bi2201 from underdoped to the heavily overdoped non-superconducting samples. Fig. S11(d-f) shows the band structures of the optimally-doped and overdoped Bi2212, including the heavily overdoped sample with a $T_{c}$ at $\sim$35\,K. The corresponding second derivative EDCs obtained from the band structures in Fig. S11(a-f) are shown in Fig. S11h. The two coexisting mode-couplings are observed in all the measurements near the antinodal region in Bi2201 and Bi2212 with different doping levels.  \\

     \noindent{\large{\bfseries 9. Simulations of the mode-coupling in the superconducting state of Bi2212}
  
     We also carried out the simulation of the single particle spectral function in the superconducting state of Bi2212 by including electron coupling with one boson mode. The main purpose is to check on the effect of the superconducting transition on the spectral function and related mode-coupling. The simulations are performed by using the finite temperature Eliashberg equations\cite{Sandvik2004,Bok2016} and we put in the boson mode with a finite linewidth.

     The ARPES intensity is obtained by multiplying the spectral weight, $A(\mathbf{k},\omega)$, by the Fermi-Dirac distribution function, $f(\omega,T)$.
    \begin{equation}
     	I(\mathbf{k},\omega,T) = f(\omega,T)A(\mathbf{k},\omega)
     \end{equation}
    \noindent where
      \begin{equation}
 	A(\mathbf{k},\omega)=-\frac{1}{\pi}Im\frac{\omega-\Sigma(\mathbf{k},\omega)+\xi(\mathbf{k})+\chi(\mathbf{k},\omega)}{(\omega-\Sigma(\mathbf{k},\omega))^2-(\xi(\mathbf{k})+\chi(\mathbf{k},\omega))^2-\phi^2(\mathbf{k},\omega)}
    \end{equation}
    The self-energies, $\Sigma$ and $\chi$, can be obtained by self-consistent Eliashberg equations:
     \begin{equation}
	\Sigma(\mathbf{k},\omega)=\sum_{\mathbf{k'}}\int_{-\infty}^{\infty}d\varepsilon\int_{-\infty}^{\infty}d\varepsilon' K(\varepsilon,\varepsilon',\omega)A_{N}(\mathbf{k'},\varepsilon)\alpha^{2}F(\mathbf{k-k'},\varepsilon')
	\label{Eliashberg}
   \end{equation}
   \begin{equation}
	\chi(\mathbf{k},\omega)=\sum_{\mathbf{k'}}\int_{-\infty}^{\infty}d\varepsilon\int_{-\infty}^{\infty}d\varepsilon' K(\varepsilon,\varepsilon',\omega)A_{\chi}(\mathbf{k'},\varepsilon)\alpha^{2}F(\mathbf{k-k'},\varepsilon')
	\label{Eliashberg}
    \end{equation}
    \noindent where $\alpha^{2}F(\mathbf{k-k'},\varepsilon')$ is the Eliashberg function and
     \begin{equation}
	A_{N}=-\frac{1}{\pi}Im\frac{\omega-\Sigma(\mathbf{k},\omega)}{(\omega-\Sigma(\mathbf{k},\omega))^2-(\xi(\mathbf{k})+\chi(\mathbf{k},\omega))^2-\phi^2(\mathbf{k},\omega)}
   \end{equation}
   \begin{equation}
	A_{\chi}=-\frac{1}{\pi}Im\frac{\xi(\mathbf{k})+\chi(\mathbf{k},\omega)}{(\omega-\Sigma(\mathbf{k},\omega))^2-(\xi(\mathbf{k})+\chi(\mathbf{k},\omega))^2-\phi^2(\mathbf{k},\omega)}
    \end{equation}
    \begin{equation}
	K(\varepsilon,\varepsilon',\omega)=\frac{f(\varepsilon)+n(-\varepsilon')}{\varepsilon+\varepsilon'-\omega-i\delta}
    \end{equation}
     \noindent Here $n(-\varepsilon')$ is Boson-Einstein distribution function. We do not leave out the particle-hole asymmetric part of the self-energy, $\chi$, for more realistic calculation.
    We simply assumed a $d$-wave superconducting gap:
   \begin{equation}
   	\Delta(\mathbf{k})=\Delta_0(\mbox{cos}(k_x)-\mbox{cos}(k_y))/2
   \end{equation}
    \noindent with $\Delta_0$ being the maximum gap at the antinodal region. We use $\Delta_0$=20\,meV. The pairing self-energy, $\phi$, can then be obtained by:
   \begin{equation}
	\phi(\mathbf{k},\omega) = \Delta(\mathbf{k})Z(\mathbf{k},\omega) = \Delta(\mathbf{k})(1-\Sigma(\mathbf{k},\omega)/\omega)
   \end{equation}
    
    In the calculation, we choose the Eliashberg function as: 
     \begin{equation}
	\alpha^{2}F(\mathbf{q},\omega)=\left(\alpha_0 (\mbox{cos}^2(\frac{q_x}{2})+\mbox{cos}^2(\frac{q_y}{2}))\right)\left( \frac{(\Gamma/2)^{2}}{(\omega-\omega_{b})^2+(\Gamma/2)^2}-\frac{(\Gamma/2)^{2}}{(\omega+\omega_{b})^2+(\Gamma/2)^2} \right)
    \end{equation}
    \noindent Here we separate the energy dependence and momentum dependence part of the Eliashberg function. It has a single peak at $\omega_{b}$=65\,meV and its width can be controlled by $\Gamma$. We use $\Gamma$=20\,meV. For the momentum dependence of the Eliashberg function, we adopt Cu-O-Cu buckling-like mode.
     
     We set the value of the $\lambda$ defined as below to be 0.5 by adjusting $\alpha_0$.
    \begin{equation}
	\lambda=\int_{-\infty}^{\infty}d\omega \int d\mathbf{q} N(\omega)\frac{\alpha^{2}F(\mathbf{q},\omega)}{\omega}
   \end{equation}
     \noindent where $N(\omega)$ is the density of state.

     The bare dispersion, $\xi(\mathbf{k})$, is given by:
    \begin{equation}
	\xi(\mathbf{k})=-2 t_1 (\mbox{cos}(k_x)+\mbox{cos}(k_y))+ 4 t_2 \mbox{cos}(k_x) \mbox{cos}(k_y) - 2 t_3 (\mbox{cos}(2k_x) + \mbox{cos}(2k_y))+\mu 
    \end{equation}
   \noindent where $t_1=0.395$, $t_2=0.084$, $t_3=0.042$ and $\mu=0.474$. Here, we set the saddle point energy to be 30\,meV. The self-energy effect lifts the saddle point energy by 10\,meV so that we have 20\,meV of the renormalized saddle point energy.
    
     Figure S12 shows the simulated spectral functions of the electron-boson coupling in Bi2212 along the nodal direction and the related energy scales. We have carried out the simulations for three cases: 100\,K in the normal state (Fig. S12(a-c)), 1\,K in the normal state (Fig. S12(d-f)) and 1\,K in the superconducting state (Fig. S12(g-i)). Fig. S12j shows the band dispersions obtained by fitting MDCs at different energies from the band structure in Fig. S12a, Fig. S12d and Fig. S12g, where we can observe clearly a kink structure in each dispersion. Fig. S12k shows the effective real part of the electron self-energies obtained from Fig. S12j. We can see that the kink structure keeps at the same position of 65\,meV for the 100\,K and 1\,K cases in the normal state. But it shifts to a higher energy of $\sim$76\,meV for the 1\,K case in the superconducting state. In the second derivative EDCs in Fig. S12l, the hump features are complicated by the van Hove singularity. The energy scale for the 1\,K case in the normal state, defined by the middle point between the dip (ND, normal state-dip) and the hump (NH, normal state-hump) features, lies at $\sim$65\,meV. For the 1\,K case in the superconducting state, the energy scale shifts to $\sim$81\,meV. These results indicate that the energy scale shifts to a higher energy when the sample enters into the superconducting state.
     
     Figure S13 shows the simulated spectral functions of the electron-boson coupling in Bi2212 along the antinodal direction and the related energy scales. Here the simulations are carried out also for three cases: 100\,K in the normal state (Fig. S13(a-c)), 1\,K in the normal state (Fig. S13(d-f)) and 1\,K in the superconducting state (Fig. S13(g-i)). In the second derivative EDCs in Fig. S13j, the energy scale for the 1\,K case in the normal state lies at $\sim$65\,meV. For the 1\,K case in the superconducting state, the energy scale shifts to $\sim$81\,meV. These results also indicate that the energy scale shifts to a higher energy when the sample enters into the superconducting state. \\

     \noindent{\large{\bfseries 10. Ubiquitous ``peak-double dip-double hump'' structure in Bi2212 and Bi2201}
     
     Figure S14 shows the identification of the peak-dip-hump structure in EDCs at the nodal Fermi momentum in Bi2201 and Bi2212. Fig. S14a shows the band structure of the Bi2212 OD73K sample taken at 17\,K along the nodal direction. We can observe two dip and two hump structures in the difference (purple line in Fig. S14c) between the EDC at17\,K (black line in Fig. S14c) obtained from the band structure of the Bi2212 OD73K sample and a fitted Lorentzian (red line in Fig. S14c). We can also observe two dip and two hump structures in the difference (purple line in Fig. S14d) between the EDC at 15\,K (black line in Fig. S14d) obtained from the band structure of the heavily overdoped non-superconducting Bi2201 sample and a fitted Lorentzian (red line in Fig. S14d). 
     
     Figure S15 shows the EDCs at different momentum points, at different temperatures, in Bi2212 and Bi2201 samples with different doping levels. The ``peak-double dip-double hump'' structure can be commonly observed in these EDCs. \\

    \vspace{3mm}
	

	\newpage

	\renewcommand{\thefigure}{S\arabic{figure}}
	
	\begin{figure*}[tbp]
	\begin{center}
			\includegraphics[width=1.0\columnwidth,angle=0]{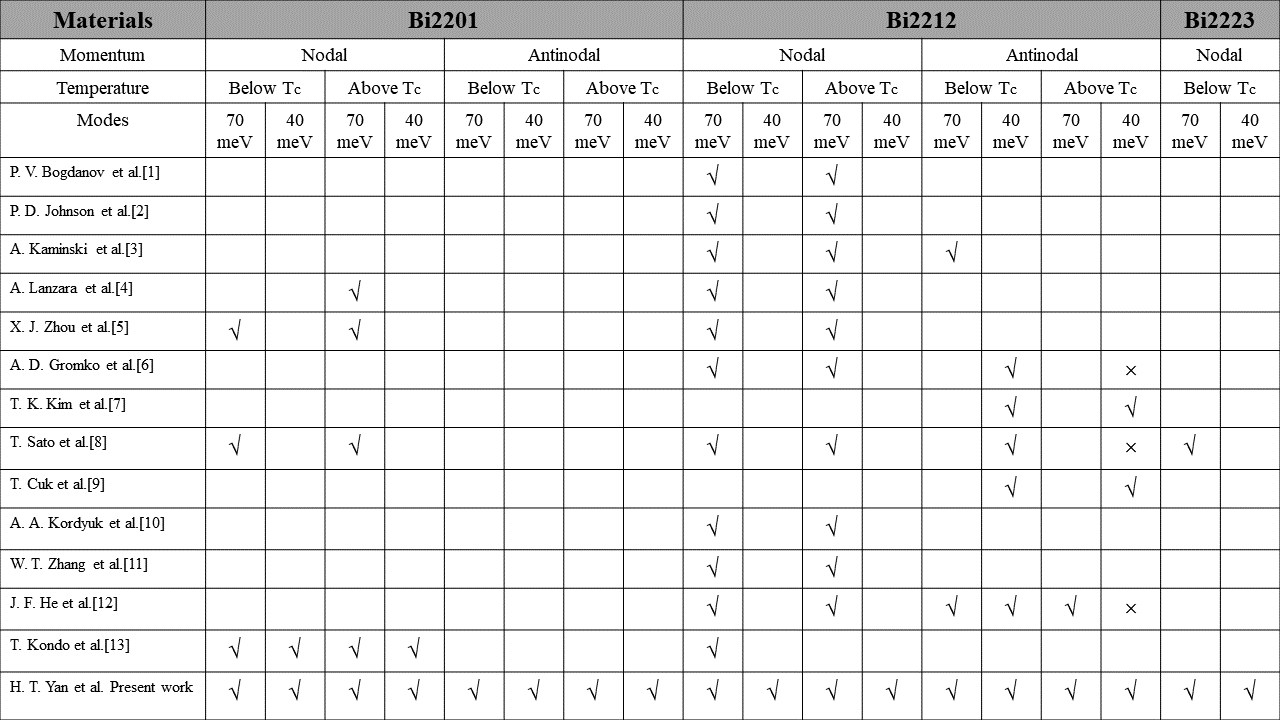}
	\end{center}
		
	\caption{{\bf Summarized ARPES observations of the electron-boson couplings in B2201, Bi2212 and Bi2223.} ``$\surd$'' represents ``observed'' while ``$\times$'' represents ``not observed''.}
		
	\end{figure*}
	
	\begin{figure*}[tbp]
    \begin{center}
		\includegraphics[width=1.0\columnwidth,angle=0]{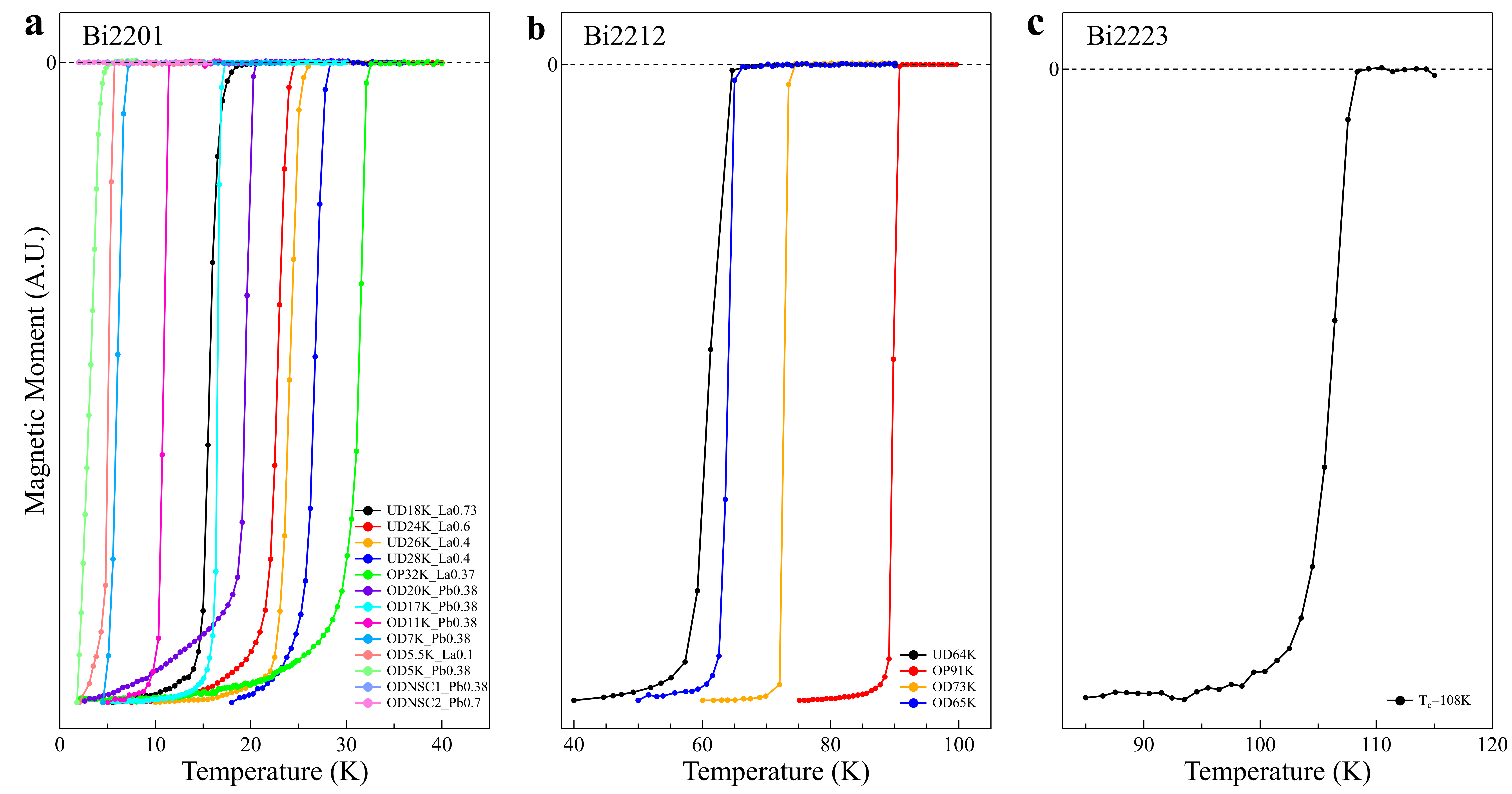}
	\end{center}
	
	\caption{{\bf Magnetic measurements of Bi2201, Bi2212 and Bi2223 samples.} (a) Magnetic measurements of Bi2201 single crystals including La-Bi2201 and Pb-Bi2201. All the Bi2201 samples are labeled by the nominal composition of the La or Pb content. (b) Magnetic measurements of Bi2212 single crystals. (c) Magnetic measurement of the Bi2223 single crystal.}
	
	\end{figure*}

    \begin{figure*}[tbp]
	\begin{center}
		\includegraphics[width=1.0\columnwidth,angle=0]{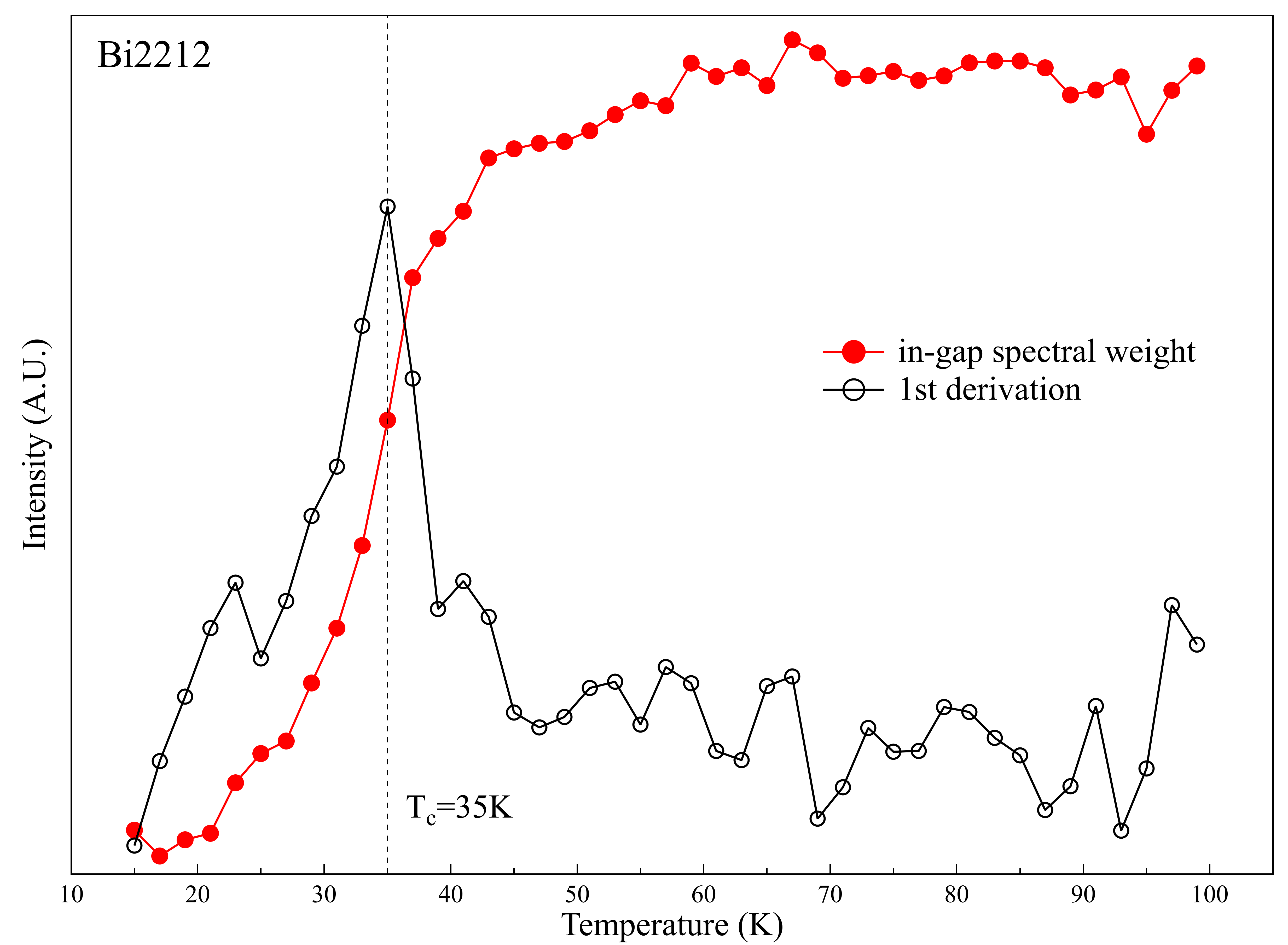}
	\end{center}
	
	\caption{{\bf Estimation of  $T_{c}$ of a Bi2212 sample annealed $\mathit{in\,situ}$ in ozone atmosphere.} The temperature dependent in-gap spectral weight (integrated within 4\,meV energy window of $E_{F}$) obtained from EDCs measured near the antinodal region at different temperatures is shown as the red dotted line. Its temperature derivative is shown as the black dotted line. $T_{c}$ is estimated to $\sim$35\,K from the largest slope in the spectral weight and the maximum in the first derivative curve\cite{ZXShen2018YHe}.}	
	\end{figure*}

	\begin{figure*}[tbp]
	\begin{center}
		\includegraphics[width=1.0\columnwidth,angle=0]{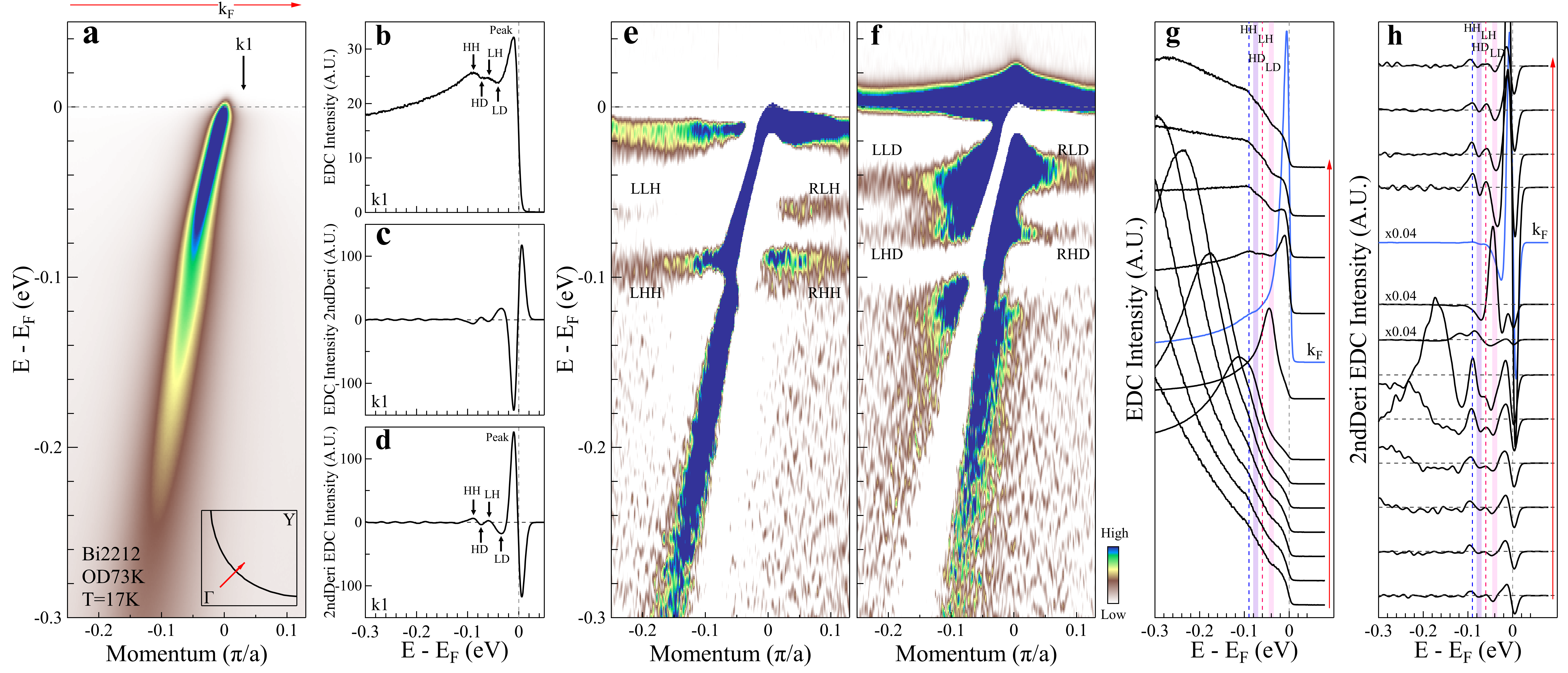}
	\end{center}
	
	\caption{{\bf Observation of two coexisting mode-couplings in the band structure measured along the nodal direction in Bi2212.} (a) Band structure of the Bi2212 OD73K sample taken at 17\,K along the nodal direction. The location of the momentum cut is shown in the bottom-right inset by the red line. (b) A typical EDC from (a) at the momentum point k1 with a main peak (peak), a low energy hump (LH), a high energy hump (HH), a low energy dip (LD) and a high energy dip (HD). (c) Second derivative of the EDC in (b). (d) Negative of the EDC second derivative in (c). It shows a good correspondence to the original EDC in (b). The peak, hump and dip structures become more pronounced. (e) Second derivative image of (a) with respect to energy showing the absolute magnitude of the negative values of the second derivative EDC intensity in (c). It highlights the peak and hump structures in the original data in (a). In addition to the main band, four features are observed and marked as LLH, RLH, LHH and RHH. (f) Second derivative image of (a) with respect to energy showing the absolute magnitude of the positive values of the second derivative EDC intensity in (c). It highlights the dip structures in the original data in (a). Four features are observed marked as LLD, RLD, LHD and RHD. (g) Typical EDCs at different momentum points in (a). For clarity, the EDCs are offset along the vertical axis. The vertical red line with an arrow in (g) corresponds to the momentum region represented by the top horizontal red line with an arrow in (a). The EDC at the Fermi momentum ($k_{F}$) is marked by blue color. Two dip features are observed marked as HD and LD and two hump features are observed marked as HH and LH. (h) Corresponding second derivative EDCs obtained from (g). For clarity, the data are offset along the vertical axis. Two dip features are marked as HD and LD and two hump features are marked as HH and LH.}
		
	\end{figure*}

	\begin{figure*}[tbp]
	\begin{center}
		\includegraphics[width=1.0\columnwidth,angle=0]{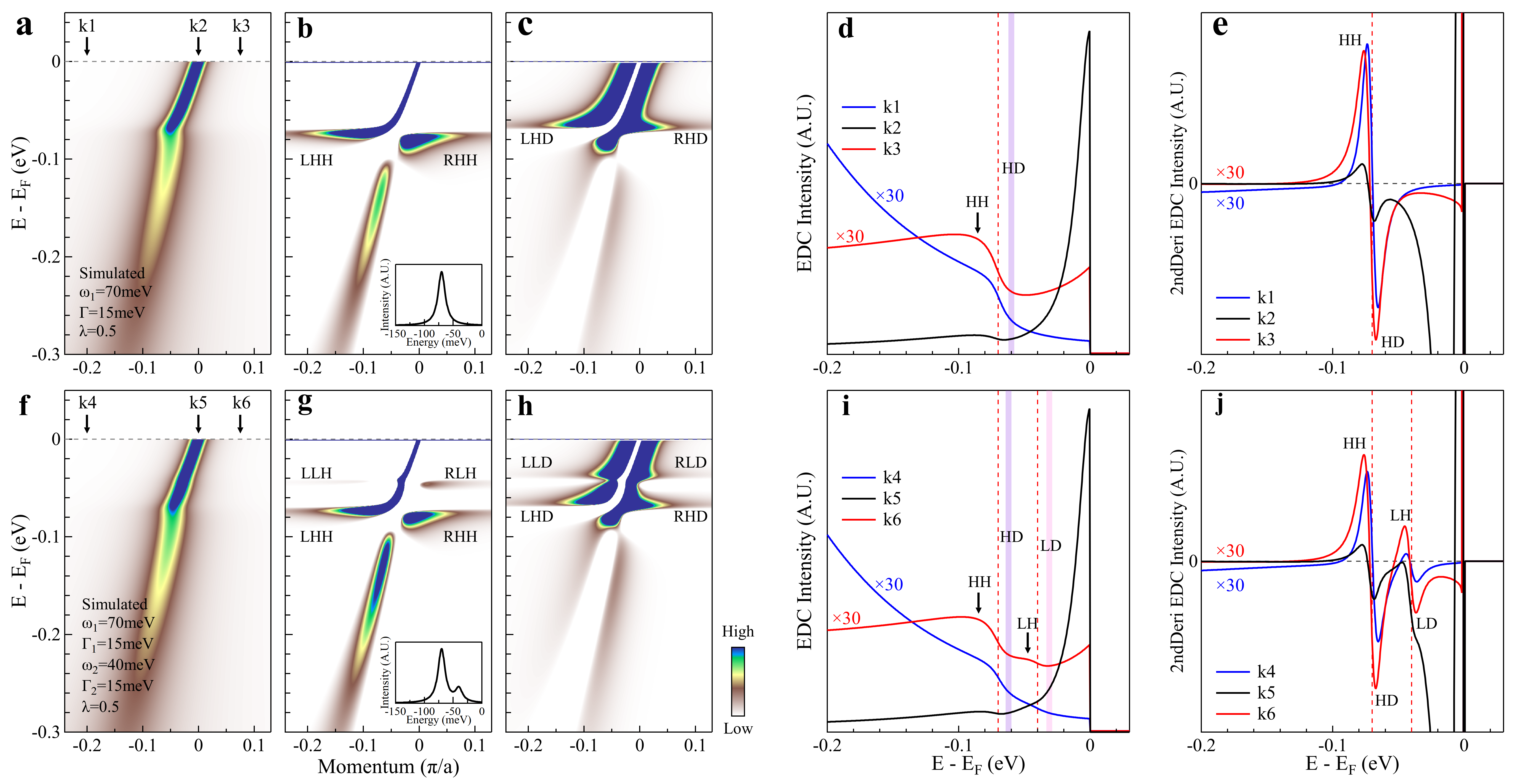}
	\end{center}
	
	\caption{{\bf The peak-dip-hump structure in the simulated spectral function of the mode-couplings involving one mode and two modes.} (a) Simulated spectral function of the mode-coupling involving one mode with an energy of 70\,meV. The mode assumed in the simulation is shown in the bottom-right inset in (b). We take a straight line as the bare band that is similar to the calculated band of Bi2212 along the nodal direction in order to make a comparison with the measured results. (b) Second derivative image of (a) with respect to energy showing the absolute magnitude of the negative values. In addition to the main band, two features appear marked as LHH and RHH. (c) Second derivative image of (a) with respect to energy showing the absolute magnitude of the positive values. Two features are observed marked as LHD and RHD. (d) EDCs from the spectral function in (a) at three momentum points, k1, k2 and k3. In addition to the EDC peaks near the Fermi level, on their left side, one dip (HD) and one hump (HH) features can be observed. The red vertical dashed line indicates the position of the mode energy at 70\,meV. (e) Second derivative EDCs obtained from (d). The HD feature in (e) corresponds to the dip structure in (d) and (LHD, RHD) features in (c). The HH feature in (e) corresponds to the hump structure in (d) and (LHH, RHH) features in (b). The red vertical dashed line indicates the position of the mode energy at 70\,meV. (f-j) Same as (a-e) but for the simulated spectral function involving two modes with energies of 40\,meV and 70\,meV. Different from the one mode-coupling case where only one dip and one hump features are produced, in the two mode-coupling case, in addition to the EDC peaks near the Fermi level, on their left side, two dip features, HD and LD, and two hump features, HH and LH, can be observed in (i). Correspondingly, two dip structures can be observed in (j) (LD and HD) and in (h) (LLD, RLD, LHD and RHD) while two hump structures can be observed in (j) (LH and HH) and in (g) (LLH, RLH, LHH and RHH).}	
	
    \end{figure*}

	\begin{figure*}[tbp]
	\begin{center}
		\includegraphics[width=1.0\columnwidth,angle=0]{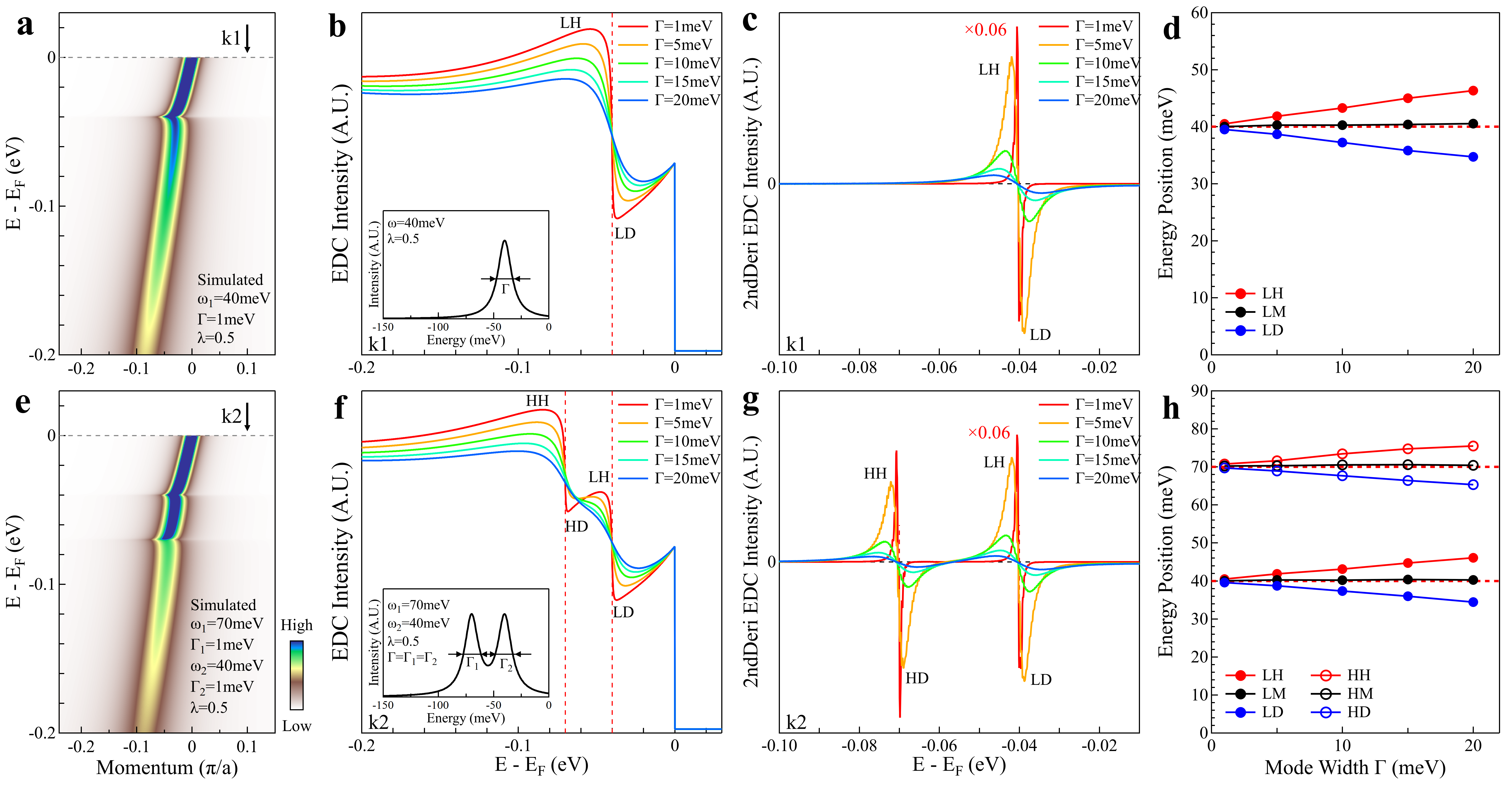}
	\end{center}
	
	\caption{{\bf Determination of the energy scales from the hump and dip structures in the simulated spectral function with different mode widths.} (a) Simulated spectral function of the mode-coupling involving one mode with an energy of 40\,meV and a FWHM of 1\,meV. (b) EDCs from the spectral functions simulated with different mode linewidths at the momentum point k1 shown in (a). The bottom-left inset shows the mode used in the simulations with the linewidth marked. A dip (LD) and a hump (LH) features can be observed in these EDCs. The red vertical dashed line indicates the position of the mode energy at 40\,meV. (c) Second derivative EDCs obtained from (b). The LD feature corresponds to the dip structure in (b) and the LH feature corresponds to the hump structure in (b). The red vertical dashed line indicates the position of the mode energy at 40\,meV. (d) The effect of the mode linewidth on the energy position of the dip feature (blue circles), hump feature (red circles) and the middle point between the dip and the hump features (black circles) obtained from the second derivative EDCs in (c). The horizontal dashed line marks the position of the mode energy at 40\,meV. (e-h) Same as (a-d) but for the simulated spectral function involving two modes with energies of 40\,meV and 70\,meV. The two horizontal dashed lines mark the positions of the mode energies at 40\,meV and 70\,meV.}
    \end{figure*}
		
	\begin{figure*}[tbp]
	\begin{center}
		\includegraphics[width=1.0\columnwidth,angle=0]{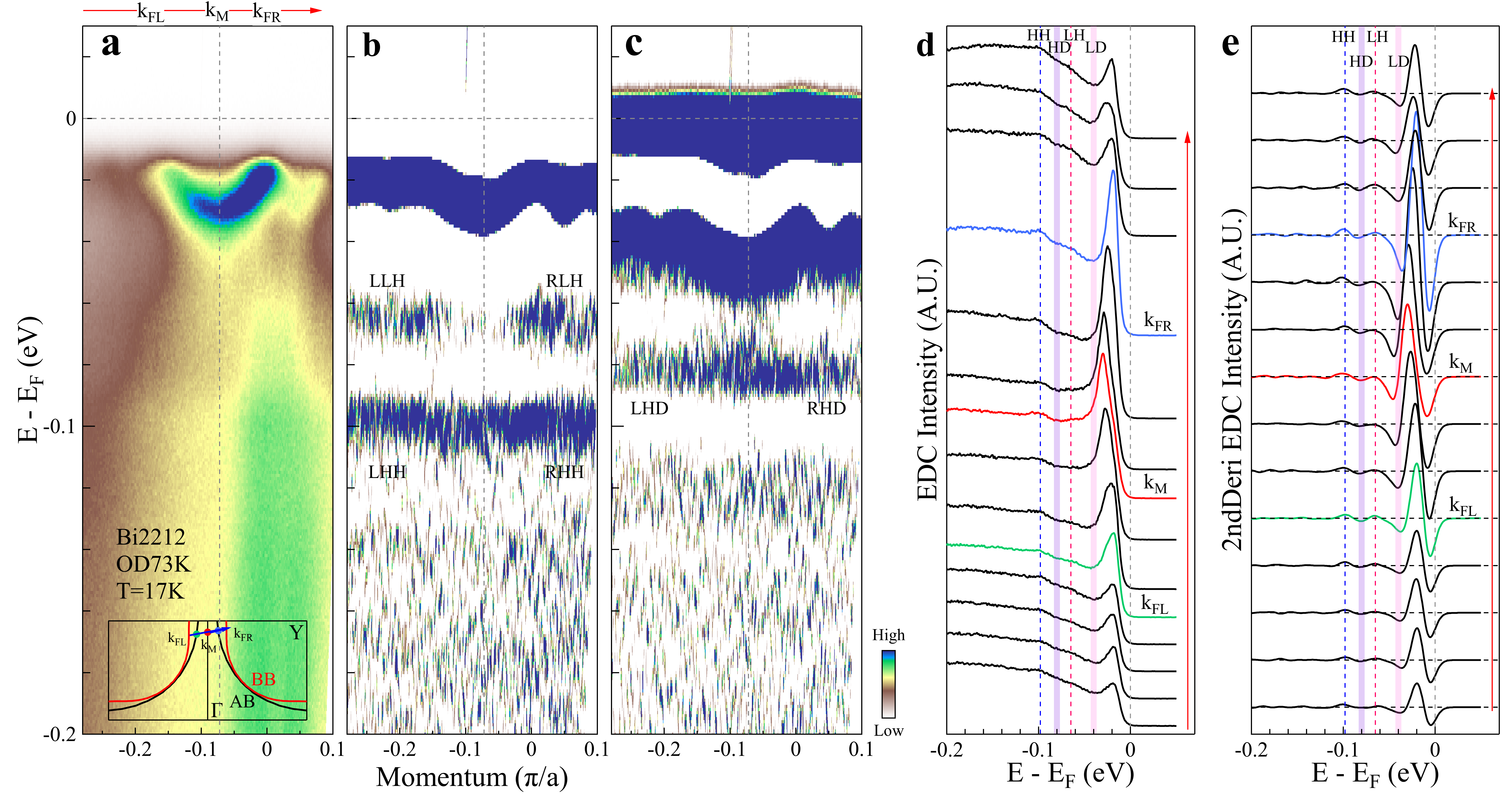}
	\end{center}
	
	\caption{{\bf Observation of two coexisting mode-couplings in the band structure near the antinodal region in Bi2212.} (a) Band structure of the Bi2212 OD73K sample taken at 17\,K near the antinodal region. The location of the momentum cut is shown in the bottom inset by the blue line where the bonding (BB) and antibonding (AB) Fermi surface sheets are both depicted. (b) Second derivative image of (a) with respect to energy showing the absolute magnitude of the negative values. In addition to the main band, four features are observed and marked as LLH, RLH, LHH and RHH with the LLH and RHH forming nearly a continuous flat feature. (c) Second derivative image of (a) with respect to energy showing the absolute magnitude of the positive values. Two features are observed marked as LHD and RHD that form nearly a continuous flat feature. (d) Typical EDCs at different momentum points in (a). For clarity, the EDCs are offset along the vertical axis. The vertical red line with an arrow in (d) corresponds to the momentum region represented by the top horizontal red line with an arrow in (a). The EDCs at three representative momentum points, $\mathrm{k_{FL}}$ (left Fermi momentum), $\mathrm{k_{M}}$ (point at the (0, 0)-$(0, \pi)$ line) and $\mathrm{k_{FR}}$ (right Fermi momentum) are marked by green, red and blue colors, respectively. The location of these three momentum points is indicated by the green, red and blue circles in the inset of (a). Two dip features are observed marked as HD and LD, and two hump features can be seen marked as HH and LH. (e) Corresponding second derivative EDCs obtained from (d). For clarity, the data are offset along the vertical axis. Two dip features are marked as HD and LD, and two hump features are marked as HH and LH.}	
	
	\end{figure*}

	\begin{figure*}[tbp]
		\begin{center}
			\includegraphics[width=1.0\columnwidth,angle=0]{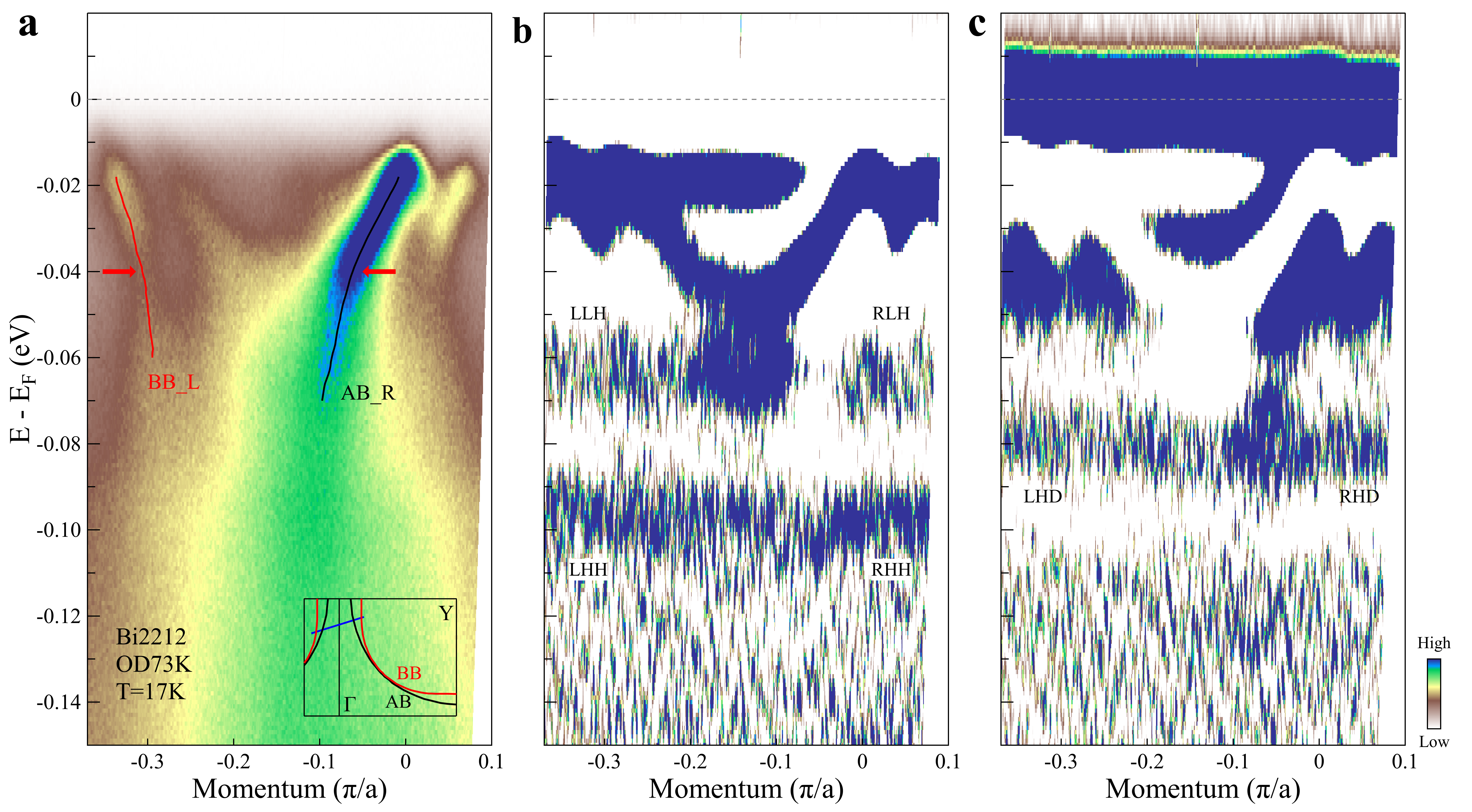}
		\end{center}
		
	\caption{{\bf Observation of the $\sim$40\,meV kink in the band structure measured at 17\,K near the antinodal region in Bi2212.} (a) Band structure of the Bi2212 OD73K sample taken at 17\,K near the antinodal region. The location of the momentum cut is shown in the bottom-right inset by the blue line where the bonding (BB) and antibonding (AB) Fermi surface sheets are both depicted. The dispersions obtained by fitting MDCs at different energies are shown as a black line for the right antibonding band  (AB$\_$R) and a red line for the left bonding band (BB$\_$L). The red arrows mark the kink positions of the AB$\_$R and BB$\_$L bands at $\sim$40\,meV. (b) Second derivative image of (a) with respect to energy showing the absolute magnitude of the negative values. In addition to the main bands, four features are observed and marked as LLH, RLH, LHH and RHH. (c) Second derivative image of (a) with respect to energy showing the absolute magnitude of the positive values. Two features are observed marked as LHD and RHD.}
	\end{figure*}

	\begin{figure*}[tbp]
	\begin{center}
		\includegraphics[width=1.0\columnwidth,angle=0]{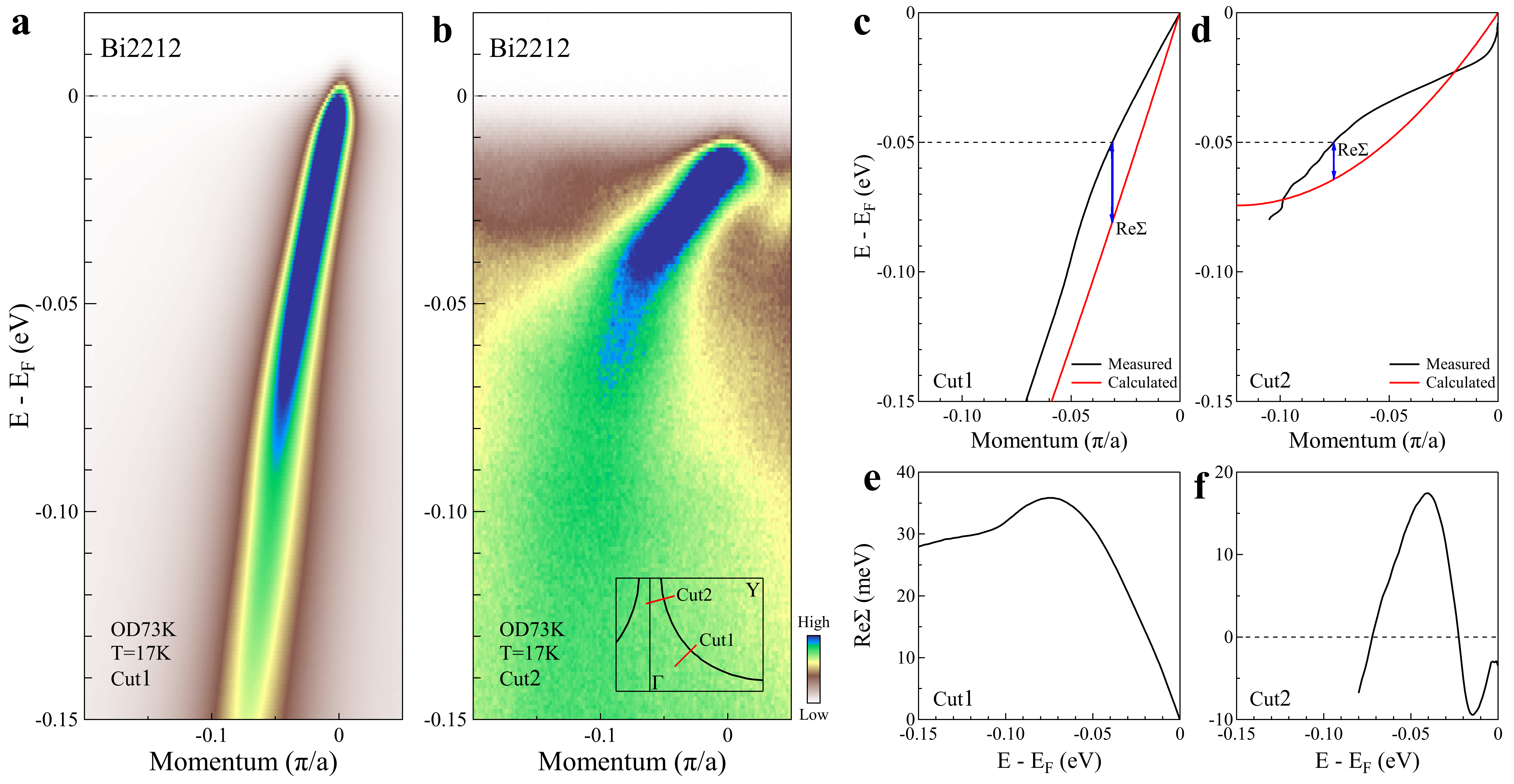}
	\end{center}
	
	\caption{{\bf Extraction of the real part of the electron self-energy.} (a) Band structure of the Bi2212 OD73K sample measured along the nodal direction at 17\,K. The location of the momentum cut is show in the bottom-right inset in (b) by a red line marked as Cut 1. (b) Same as (a) but for the momentum Cut 2 near the antinodal region. (c) Band dispersion (black line) obtained by fitting MDCs at different energies from (a). The red line represents the band dispersion from the band structure calculations\cite{ABansil2005RSMarkiewicz}. It is used as the bare band to extract the real part of the electron self-energy. (d) Same as (c) but obtained from the band structure in (b). (e) The real part of the electron self-energy from the dispersion in (c). (f) Same as (e) but from the dispersion in (d).} 	
	 	
	\end{figure*}

    \begin{figure*}[tbp]
	\begin{center}
		\includegraphics[width=1.0\columnwidth,angle=0]{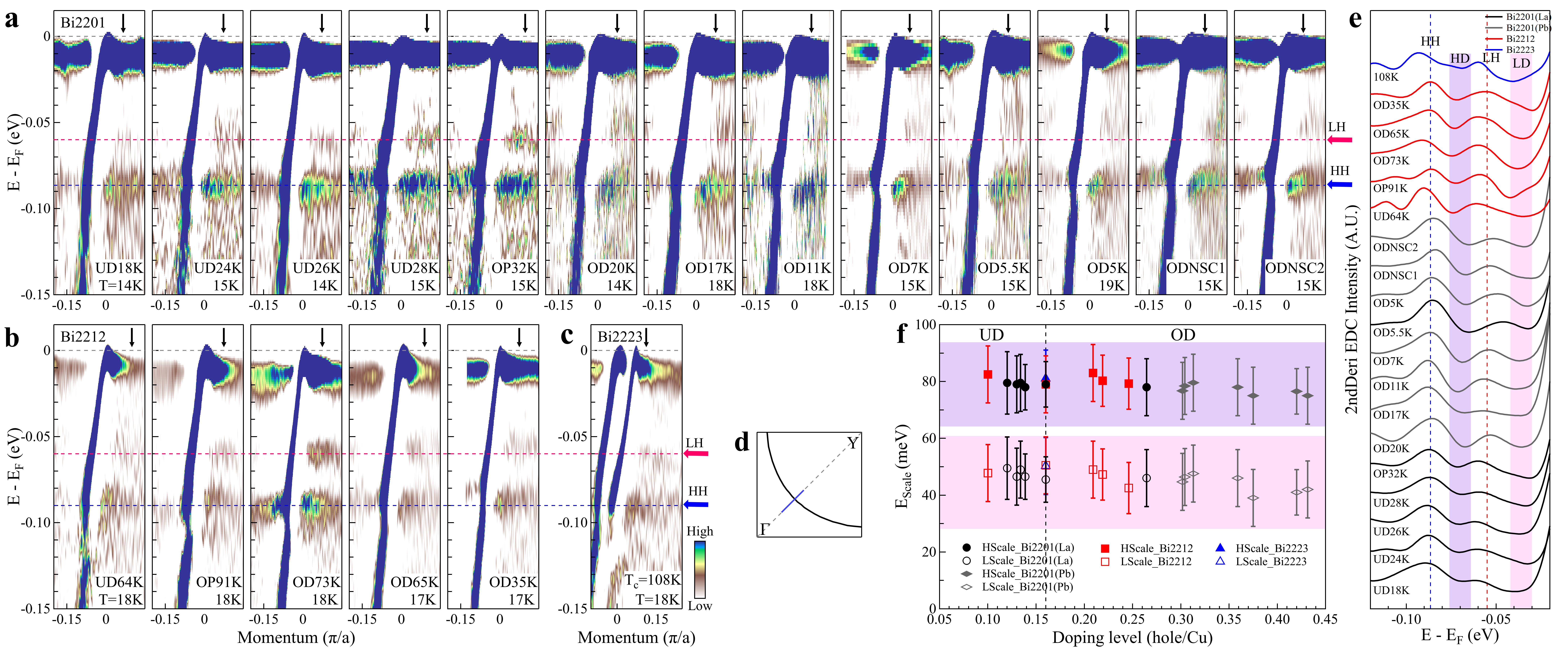}
	\end{center}
	
	\caption{{\bf Doping dependence of the two mode-couplings in Bi2201, Bi2212 and Bi2223 along the nodal direction.} (a) Band structures of Bi2201 with different doping levels taken along the nodal direction at 14$\sim$19\,K. The location of the momentum cut is shown in (d) by a blue line. These are the second derivative images with respect to energy showing the absolute magnitude of the negative values. (b) Same as (a) but for Bi2212 with different doping levels measured at 17$\sim$18\,K. (c) Same as (a) but for the Bi2223 sample with a $T_{c}$ of 108\,K measured at 18\,K. The two features observed in (a, b and c) are marked by LH (pink dashed line) and HH (blue dashed line). (e) Second derivative EDCs obtained from the band structures in (a, b and c). The corresponding momentum positions are shown by arrows in each panel of (a, b and c). The two hump features are labeled as HH and LH, and the two dip features are labeled as HD and LD. For clarity, the data are normalized by the intensity difference between the HH and LD features, and are offset along the vertical axis. (f) Summarized energy positions of the two energy scales obtained from (e) for Bi2201 (black and gray symbols), Bi2212 (red symbols) and Bi2223 (blue symbols). The energy position of the high energy mode (HMode) is determined from the middle point between the HH and HD features while the position of the low energy mode (LMode) is from the middle point between the LH and LD features in (e).}
		
    \end{figure*}

	\begin{figure*}[tbp]
	\begin{center}
		\includegraphics[width=1.0\columnwidth,angle=0]{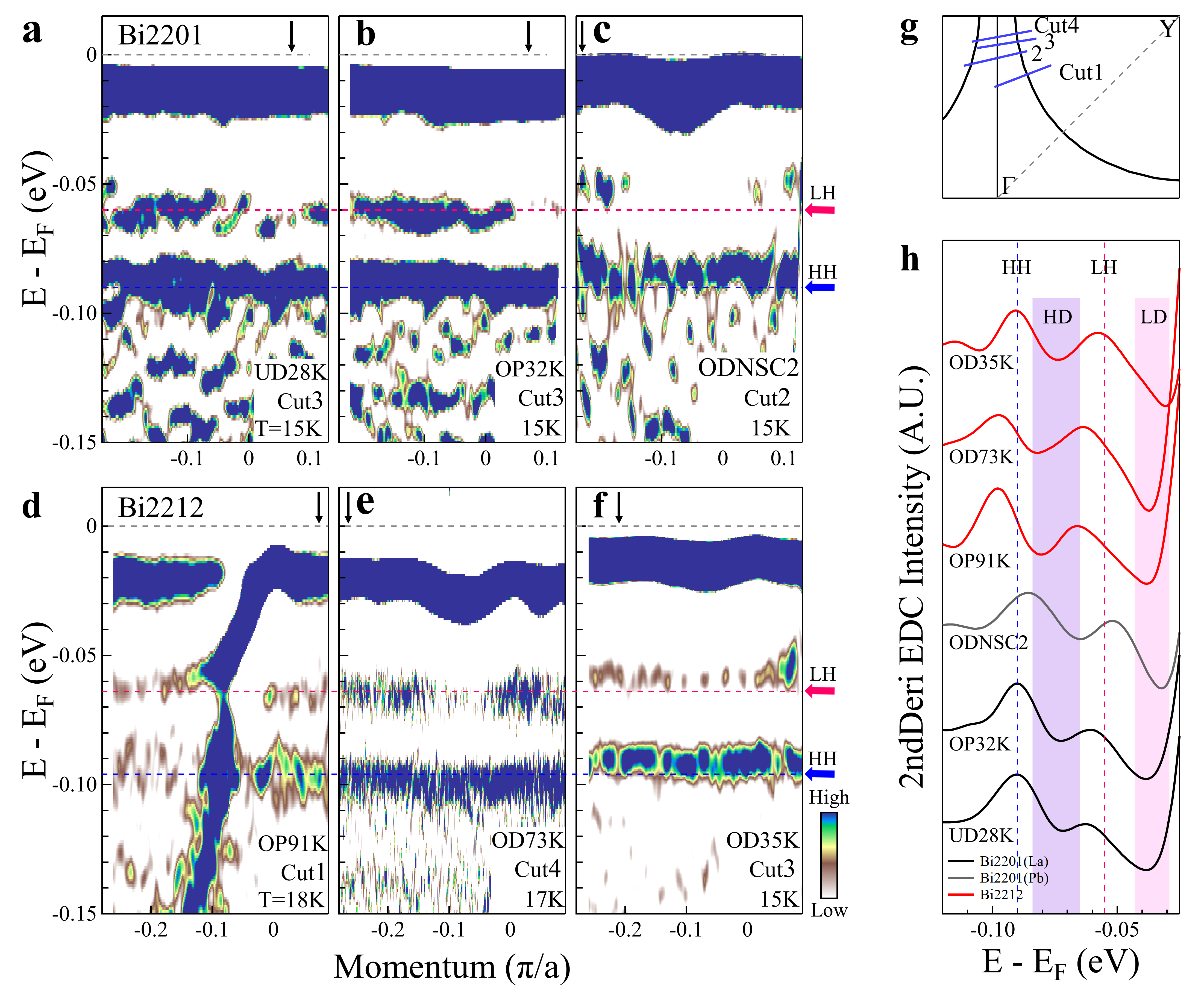}
	\end{center}
	
	\caption{{\bf Doping dependence of the two mode-couplings in Bi2201 and Bi2212 near the antinodal region.} (a-c) Band structures of Bi2201 with different doping levels, including underdoped (a, Cut3), optimally-doped (b, Cut3) and overdoped (c, Cut2), taken along the momentum cuts near the antinodal region at 15\,K. The locations of the momentum cuts are shown in (g) by the blue lines. These are the second derivative images with respect to energy showing the absolute magnitude of the negative values. (d-f) Same as (a-c) but for Bi2212 with different doping levels along the momentum cuts near the antinodal region, including optimally-doped (d, Cut 1), overdoped (e, Cut 4) and heavily overdoped (f, Cut 3), measured at 15$\sim$18\,K. The two features observed in (a-f) are labeled as LH (pink dashed line) and HH (blue dashed line). (g) The location of the momentum cuts used in (a-f). (h) Second derivative EDCs obtained from the band structures in (a-f). The corresponding momentum positions are shown by arrows in each pane of (a-f). The two hump features are marked by HH and LH, and the two dip features are marked by HD and LD. For clarity, the data are normalized by the intensity difference between the HH and LD features, and are offset along the vertical axis.}
	
    \end{figure*}

	\begin{figure*}[tbp]
	\begin{center}
		\includegraphics[width=1.0\columnwidth,angle=0]{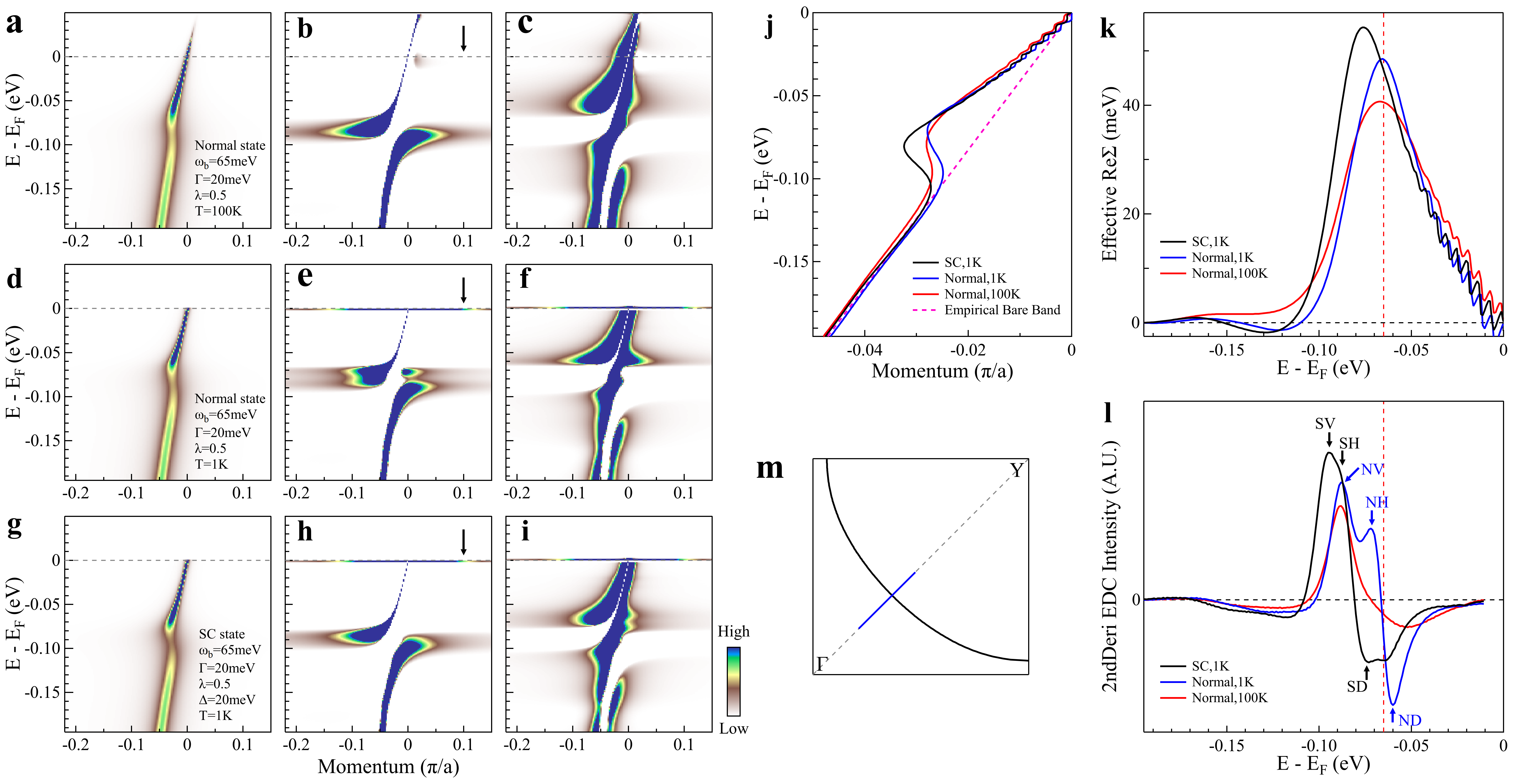}
	\end{center}
	
	\caption{{\bf Simulated spectral functions of the electron-boson coupling in Bi2212 along the nodal direction and related energy scales.} The simulation involves one mode with an energy of 65\,meV and a linewidth of 20\,meV. The location of the involved momentum cut is shown in (m) by the blue line. (a) Simulated spectral function at 100\,K in the normal state. (b) Second derivative image of (a) with respect to energy showing the absolute magnitude of the negative values. (c) Second derivative image of (a) with respect to energy showing the absolute magnitude of the positive values. (d-f) Same as (a-c) but for the simulated spectral function at 1\,K in the normal state. (g-i) Same as (a-c) but for the simulated spectral function at 1\,K in the superconducting state with a superconducting gap of 20\,meV. (j) Band dispersions obtained by fitting MDCs at different energies from the band structures in (a), (d) and (g). The pink dashed straight line represents an empirical bare band to extract the effective real part of the electron self-energy from the dispersions. (k) Effective real part of the electron self-energy obtained from the dispersions in (j). The vertical dashed line indicates the position of the mode energy at 65\,meV. (l) Second derivative EDCs obtained from the second derivative images at the momentum point marked by the arrows in (b), (e) and (h). The vertical dashed line indicates the position of the mode energy at 65\,meV. In the normal state 1\,K curve (blue line), the dip and hump positions are marked by ND and NH, respectively, and the feature caused by the van Hove singularity is marked by NV. In the superconducting state 1\,K curve (black line), the dip and hump positions are marked by SD and SH, respectively, and the feature caused by the van Hove singularity is marked by SV.}
	
   \end{figure*}

	\begin{figure*}[tbp]
	\begin{center}
		\includegraphics[width=1.0\columnwidth,angle=0]{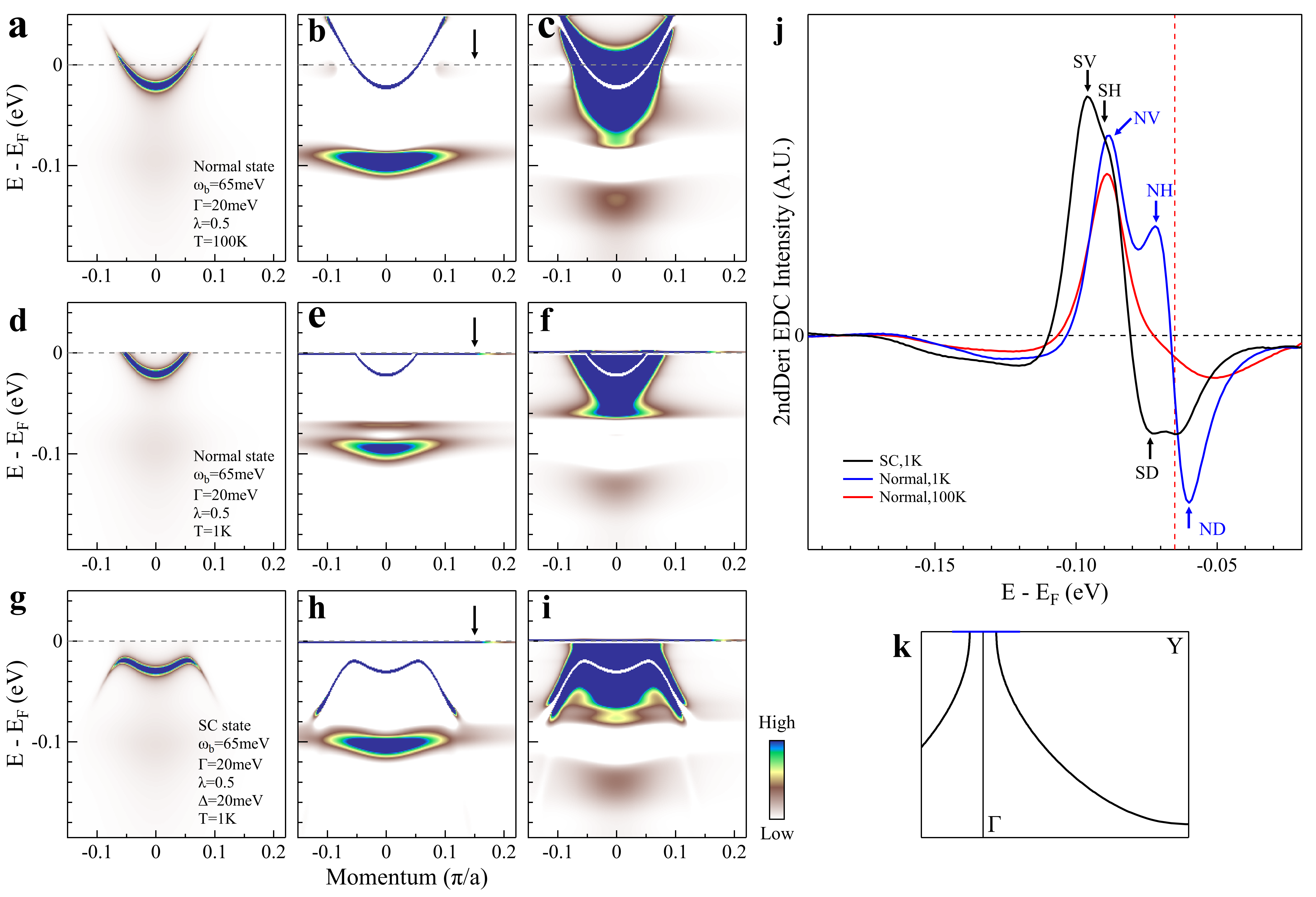}
	\end{center}
	
	\caption{{\bf Simulated spectral functions of the electron-boson coupling in Bi2212 along the antinodal direction and related energy scales.} The simulation involves one mode with an energy of 65\,meV and a linewidth of 20\,meV. The location of the involved momentum cut is shown in (k) by the blue line. (a) Simulated spectral function at 100\,K in the normal state. (b) Second derivative image of (a) with respect to energy showing the absolute magnitude of the negative values. (c) Second derivative image of (a) with respect to energy showing the absolute magnitude of the positive values. (d-f) Same as (a-c) but for the simulated spectral function at 1\,K in the normal state. (g-i) Same as (a-c) but for the simulated spectral function at 1\,K in the superconducting state with a superconducting gap of 20\,meV. (j) Second derivative EDCs obtained from the second derivative images at the momentum point marked by the arrows in (b), (e) and (h). The vertical dashed line indicates the position of the mode energy  at 65\,meV. In the normal state 1\,K curve (blue line), the dip and hump positions are marked by ND and NH, respectively, and the feature caused by the van Hove singularity is marked by NV. In the superconducting state 1\,K curve (black line), the dip and hump positions are marked by SD and SH, respectively, and the feature caused by the van Hove singularity is marked by SV.}
		
    \end{figure*}

	\begin{figure*}[tbp]
	\begin{center}
		\includegraphics[width=1.0\columnwidth,angle=0]{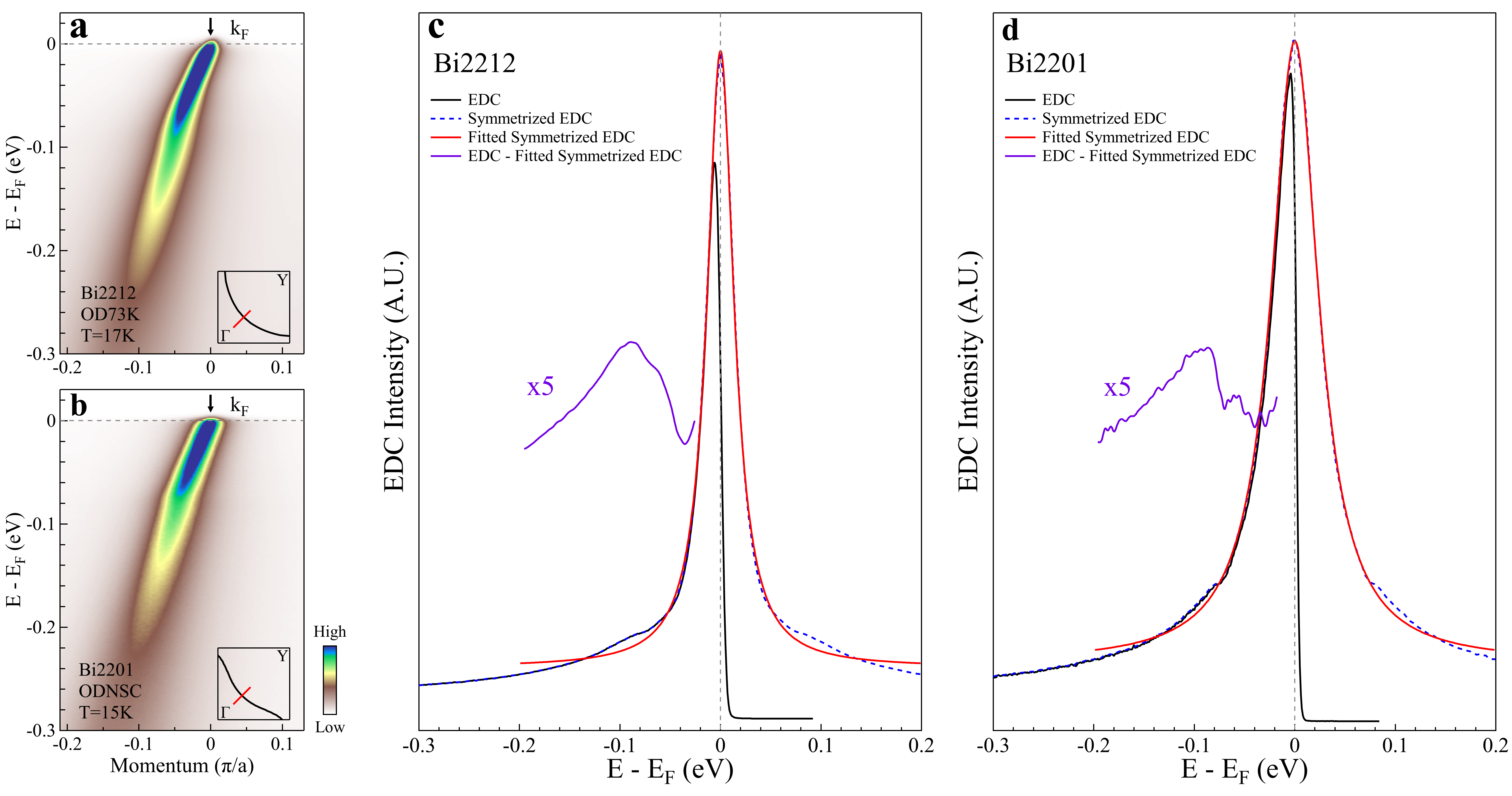}
	\end{center}
	
	\caption{{\bf Identification of the peak-dip-hump structures in EDCs at the nodal Fermi momentum in Bi2201 and Bi2212.} (a) Band structure of the Bi2212 OD73K sample taken at 17\,K along the nodal direction. The location of the momentum cut is shown in the bottom-right inset by the red line. (b) Same as (a) but for the Bi2201 ODNSC sample taken at 15\,K. (c) EDC at the Fermi momentum ($\mathrm{k_{F}}$) (black line) obtained from the image in (a). The EDC is symmetrized with respect to the Fermi level to obtain the symmetrized EDC (blue dashed line). The symmetrized EDC is then fitted by a Lorentzian (red line) in the energy range of [-0.2,0.2]\,eV. It is found that the measured EDC deviates from the fitted Lorentzian form. The difference between the measured EDC and the fitted Lorentzian in the energy range of [-0.2,-0.025]\,eV is shown by the purple solid line. For clarity, it is multiplied by five times ($\times$5). (d) Same as (c) but for the nodal EDC taken from the band structure of the Bi2201 ODNSC sample in (b).}
	
    \end{figure*}

    \begin{figure*}[tbp]
	\begin{center}
		\includegraphics[width=1.0\columnwidth,angle=0]{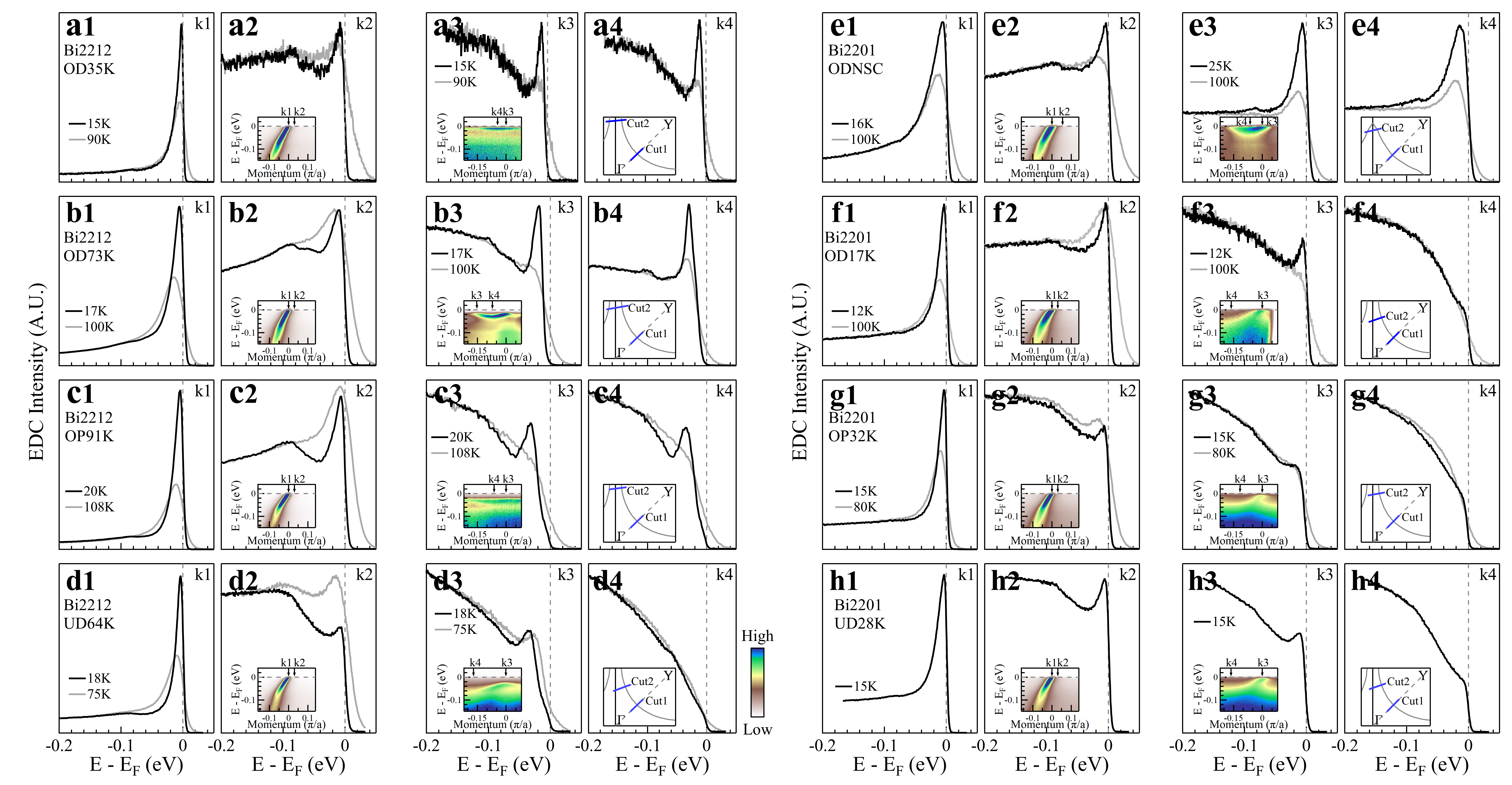}
	\end{center}
	
	\caption{{\bf Ubiquitous ``peak-double dip-double hump" structure in Bi2212 and Bi2201.} (a1-a2) EDCs from the band structure of Bi2212 OD35K sample measured along the nodal direction at two momentum points k1 (Fermi momentum, a1) and k2 (a2). The measured band structure is shown in the inset of (a2) where the location of the two momentum points k1 and k2 is marked by the arrows. The location of the momentum cut is shown in the inset of (a4) by a blue line marked as Cut 1. Two EDCs are shown in (a1) and (a2) that are measured at 15\,K (black curve) and 90\,K (gray curve). (a3-a4) Same as (a1-a2) but for the EDCs at two momentum points k3 (Fermi momentum, a3) and k4 (point at the (0, 0)-(0, $\pi$) line, a4) near the antinodal region. The measured band structure is shown in the inset of (a3) where the location of the two momentum points k3 and k4 is marked by the arrows. The location of the momentum cut is shown in the inset of (a4) by the blue line marked as Cut 2. Two EDCs are shown in (a3) and (a4) that are measured at 15\,K (black curve) and 90\,K (gray curve). (b1-b4) Same as (a1-a4) but for the Bi2212 OD73K sample measured at 17\,K and 100\,K. (c1-c4) Same as (a1-a4) but for the Bi2212 OP91K sample measured at 20\,K and 108\,K. (d1-d4) Same as (a1-a4) but the Bi2212 UD64K sample measured at 18\,K and 75\,K. (e1-e4) Same as (a1-a4) but for the Bi2201 ODNSC sample measured at 16\,K and 100\,K in (e1-e2) and at 25\,K and 100\,K in (e3-e4). (f1-f4) Same as (a1-a4) but for the Bi2201 OD17K sample measured at 12\,K and 100\,K. (g1-g4) Same as (a1-a4) but for the Bi2201 OP32K sample measured at 15\,K and 80\,K. (h1-h4) Same as (a1-a4) but for the Bi2201 UD28K sample measured at 15\,K.}
	
    \end{figure*}